\documentclass[journal,10pt,twocolumn]{IEEEtran}
\usepackage[T1]{fontenc}
\usepackage{setspace}

\ifCLASSINFOpdf
\else
\fi
\usepackage{enumerate}
\usepackage{graphicx}
\usepackage{bm}
\usepackage{caption}
\usepackage{algorithm}
\usepackage{color}
\usepackage{cite}
\usepackage{graphicx}
\usepackage{amssymb}
\usepackage{subfigure}
\usepackage{epsfig}
\usepackage{framed}
\usepackage{algorithmic}
\usepackage{multirow}
\usepackage{multicol}
\usepackage{color,xcolor}
\usepackage{amsmath}

\newtheorem{theorem}{Theorem}
\newtheorem{lem}{Lemma}

\graphicspath{{picture/}}

\usepackage[cmintegrals]{newtxmath}
\begin{document}
%
\title{Target Control of Directed Networks based on Network Flow Problems}
%
%
%
\author{Guoqi~Li, {\it  IEEE Member}, Xumin~Chen,   Pei~Tang,   Gaoxi~Xiao, {\it  IEEE Member},    Changyun Wen, {\it  IEEE Fellow} ~and~Luping~Shi
\thanks{ The work was partially supported by National Science Foundation of China (61603209), and the Study of Brain-Inspired Computing System of Tsinghua
University program (20151080467), and Ministry of Education, Singapore, under contracts  MOE2014-T2-1-028 and MOE2016-T2-1-119.}
\thanks{G. Li, P. Tang and L. Shi are with Department of Precision Instrument, Tsinghua  University, P. R. China. (e-mail: liguoqi@mail.tsinghua.edu.cn;  tangp14@mails.tsinghua.edu.cn;     lpshi@mail.tsinghua.edu.cn ).
     X. Chen is with the Department of Computer Science, Tsinghua  University, P. R. China. (e-mail:chen.xm.mu@gmail.com).
   G. Xiao and C. Wen are  with the School of EEE, Nanyang Technological University, Singapore. (e-mail: egxxiao@ntu.edu.sg, ecywen@ntu.edu.sg).
   }
\thanks{}}

\maketitle

\markboth{IEEE TRANSACTIONS ON CONTROL OF NETWORK SYSTEMS} {Shell
\MakeLowercase{\textit{et al.}}: Bare Demo of IEEEtran.cls for
Journals}
\begin{abstract}
Target control of  directed  networks,  which aims to control only  a  target   subset   instead of the  entire  set of nodes  in   large natural and technological networks,   is   an outstanding challenge faced in various  real world  applications.
We address  one  fundamental issue  regarding this   challenge,   i.e.,  for a given   target   subset,   how  to allocate  a minimum number of control sources which provide  input signals   to the network nodes.     This   issue remains     open  in general networks with loops. We show that the issue  is  essentially  a path cover problem and   can be  further converted into a maximum network flow problem.
A method termed ``Maximum Flow based  Target Path-cover''   (MFTP)  with complexity  $O(|V|^{1/2}|E|)$ in which $|V|$ and $|E|$ denote  the number of network nodes and edges   is proposed. It  is  also rigorously  proven to  provide   the minimum number of  control sources on  arbitrary directed networks,   whether  loops exist or not.  We anticipate that this  work would serve wide applications in  target  control of real-life networks,  as well as counter control of various complex  systems which may contribute to enhancing system robustness and resilience.
\end{abstract}



%
\IEEEpeerreviewmaketitle

\section{Introduction}
Over the past decade complex natural and technological
systems that permeate many aspects of daily life---including human
brain intelligence, medical science, social science, biology, and
economics---have been widely
studied~\cite{watts1998collective,strogatz2001exploring,barabasi2003scale}. Recent efforts mainly focus on the
structural controllability  of  directed  networks  \cite{liu2011controllability,ruths2014control,murota2009matrices,liu2016control}
with linear dynamics  ${\dot{\mathbf{x}}}(t)={A\mathbf x}(t)+{B\mathbf{u}}(t)$  where $\mathbf x \in R^{N \times 1}$,   $A\in R^{N \times N}$, $B\in R^{N \times M}$ and $\mathbf{u}\in R^{M \times 1}$  denote  the dimensional  system states,  adjacency matrix,  input matrix and   input signals, respectively, where $N$ denotes  the network size and $M$ is the number of external input control sources.   Maximum matching,  a classic concept in graph theory, has been  successfully  and efficiently used to  allocate the minimum number of external control sources    which provide  input signals   to the network nodes,  to guarantee the structural controllability of the entire  network \cite{liu2011controllability}\cite{commault2002characterization}.    However,
it  is often unfeasible or unnecessary to fully control the entire
large-scale networks, which motivates the control of a prescribed
subset, denoted as a target set $S$,  of large natural and technological networks.    This specific form of output
control is known  as {\it target control}.  In  \cite{klickstein2017energy}, it is claimed  that
 the required energy cost to  target control  can be reduced substantially.
Nevertheless, to   the best of our knowledge,  target control   remains largely  an outstanding challenge faced in  various real world  applications  including  the
areas of biology, chemical engineering and economic networks \cite{gao2014target}.
   Therefore, allocating a  minimum number of external control sources   to guarantee  the structural controllability of   target set $S$ instead of the  whole network, termed as  ``target controllability'' problem,     becomes  an essential issue that must be solved.


%

In \cite{gao2014target}, Gao et al.  first   considered  the target controllability  problem  and proposed a $k-$walk theory to address this problem. The  $k-$walk theory shows that when  the length of the path from  a    node  {\it denoted as  Node 1  }   to
each target node   is unique,  only  one
driver node  {\it   (Node 1)  }   is needed \cite{gao2014target}.  This interesting discovery  is,  however,   only applicable  to  directed-tree like networks with single input case. In   \cite{monshizadeh2015strong} \cite{van2017distance},    target
controllability   is also  investigated  from
a topological viewpoint  based on a constructed distance-information preserving topology matrix.
For example,  as mentioned in  \cite{van2017distance}, the  matrix $A$   can be  found using
 $p$ runs of Dijkstra's  algorithm  \cite{dijkstra1959note}.  { If  the matrix $A$  in  \cite{van2017distance}   becomes  an  adjacency matrix   as mentioned earlier  in this work,  existing schemes cannot be
applied   and  new  models  are yet to be  considered.   Thus, while the problem
of  ``structural controllability''  has been well
solved  by means of maximum-matching, and
 even  some very  interesting    algorithms   have been presented in \cite{gao2014target}\cite{monshizadeh2015strong} \cite{van2017distance} to investigate  the problem of  target controllability    in general real life networks   ${\dot{\mathbf{x}}}(t)={A\mathbf x}(t)+{B\mathbf{u}}(t)$,  target controllability for general real lift networks where the network topology described by the adjacency matrix  that usually  contains  many loops, has remained
as an open problem.  In this work, we  shall  develop  a theory to solve this problem and have shown that
the effectiveness of our method through an illustrative example and some numerical experiments on synthetic
and natural complex networks.}

 Specifically, we  address  one  fundamental issue  regarding  target control of real-life networks in this  paper, which  is  to allocate a  minimum number of control sources  for a given   target   subset $S$.
 We  show that   the  issue  is essentially    a path cover problem,  which     is to locate a set of   directed paths  denoted  as $P$  and circles  denoted  as $\mathcal{C}$  to cover $S$.   {  The   minimum number of external control sources is equal to the minimum number of directed paths in
  $P$   denoted as $|P|$ as long as $|P|\neq 0$),}   and has nothing to do with the number of  circles in  $\mathcal{C}$.   Then, we    uncover that the  path cover problem can be further
 transformed to a  maximum network  flow
problem in graph theory by building a flow network under specific  constraint conditions.   { A   ``{\it Maximum  flow based  target  path-cover}''  (MFTP)  algorithm     is presented  to obtain the  solution of the maximum flow problem.
 By  proving    the validity of such a model transformation,  the  optimality of  the  proposed MFTP is      rigorously established.
We also obtain computational complexity of  MFTP as    $O(|V|^{1/2}|E|)$, where $|V|$ and $|E|$ denote  the number of network nodes and edges, respectively.
Generally speaking,  target controllability is more difficult to be determined than the   controllability of   an  entire  network as fundamentally  it becomes a
 different  problem.
However, compared  with the computational complexity of  MM for solving the structural controllability of entire network  with  $O(|V|^{1/2} |E|)$ in \cite{liu2011controllability},  MFTP is consistent with the MM algorithm  when $S=V$.
This implies that we   can always solve the target controllability of  a target node set $S \subseteq V$ based on  MFTP. }

%


{
 There are also some rated works \cite{olshevsky2014minimal}\cite{sharma2017multitype}\cite{guo2018novel}, where some interesting aspects of network controllability are considered. For example, for the   minimal  controllability  problem  in \cite{olshevsky2014minimal}, the input matrix  $B$ is considered as $N\times N$ dimensional  diagonal matrix and
 the authors are seeking to minimize the number of nonzero entries of $B$,  and  this problem is shown to be NP-hard.
In \cite{sharma2017multitype}\cite{guo2018novel},  some applications of target controllability are  demonstrated  in   new therapeutic targets for disease intervention  in  biological networks  tough the rigorous  theoretical results are undergoing research.
 To the best of our knowledge, for the first time,   this work  provides  solutions for target control of   directed  networks   with rigorous  minimum number of control sources.
We build a link from  target   controllability to   network flow problems and anticipate that  this
would serve as the entry point leading to real  applications in target  control of   real-life complex
systems.}

\bigskip
\section{Target  Control  of Directed Networks}
\subsection{Target controllability}
We first  investigate the problem of target controllability of directed networks with  linear  dynamics using   the  minimum  number of external control sources.  Although most of real systems exhibit  nonlinear dynamics, studying their linearized dynamics is a prerequisite for studying those systems.

Without loss of generality,   define  a directed graph  $D(V,E)$  where $V$  is the  node set   $V=\{v_{1}, ... , v_{N}\}$ and $E$ is the edge set.
Denote      the  target node  set  that needs to be directly controlled  as $S=\{v_{s_1}, v_{s_2},  ...,  v_{s_{|S|}}\}$ where $s_1, ... ,s_i, ...,s_{|S|}$ are  indexes of the   nodes  in $S$ and  $|S|$ is  the cardinality of $S$, i.e., the number of nodes  in $S$.  Obviously, we have $S\subseteq V$ and $|S|\leq N$.
 In this paper,     we consider  the following linear time-invariant  (LTI) dynamic system
\begin{equation}
   \begin{array}{lll}
&{\dot{\mathbf{x}}}(t)={A\mathbf x}(t)+{B\mathbf{u}}(t) \\
&\mathbf y(t)=C \mathbf x(t)
   \end{array}
   \label{systemmodel1}
\end{equation}
where
$\mathbf{x}(t)=[x_{1}(t),\ldots, x_{N}(t)]^T$ is the state vector of $V=\{v_{1}, ... , v_{N}\}$ at time $t$ with an initial state $\mathbf{x}(0)$,
$\mathbf{u}(t)=[u_1(t), \ldots, u_{M}(t)]^T$ is the time-dependent
external control input vector of $M$ external control sources
  in which the same  control  source   input  $u_i(t)$ may  connect  to  multiple nodes, and $\mathbf y(t)=[y_1(t),  \ldots, y_{|S|}(t)]^T$  represents the output
vector  of  a target set $S$.
The matrix ${A}=[a_{ij}]_{N\times N}$ is an  adjacency matrix of
the network,  where  $a_{ij}=1$ if there is a link connecting node $i$
to node $j$;  and $a_{ij}=0$ otherwise. And the link weight denotes the connection strength. ${B}=[b_{im}]_{N\times M}$ is
 an  input matrix where $b_{im}$ is nonzero when control source $m$ is
connected to node $i$ and zero otherwise.  $C=[c_{ik}]_{|S|\times N}=[I(s_1), ... ,I(s_i), ...,I(c_{|S|})]$ is  an output matrix    where $I(s_i)$ denotes the $s_i$th row of an $N\times N$ identity matrix,   when $k=s_i$  $(i = 1, 2, ... , {|S|})$   and  $s_i$ is the $i-$th target node in $S$,  ${C}_{i{s_i}}=1$   and all
other elements are zero.

The objective is to  determine the minimum   number  (i.e. the smallest $M$) of     external control sources   which are  required to   connect  to $N$ nodes such that the state of  $S$ can be driven to any desired final state in finite time for a proper designed $\mathbf{u}(t)$. Therefore, the system $(A,B, C)$ is said to be  target controllable   \cite{gao2014target} if and only if $rank[CB, CAB, ... , CA^{N-1}B]=|S|$ for an determined input matrix $B$, a pre-given $A$ and  the chosen target node set $S$.

When both $B$ and $C$  are pre-given such that  $(A,B, C)$ is  target controllable,   we design the input signal $\mathbf{u}(t)$  as
\begin{equation}
   \begin{array}{lll}
&\mathbf{u}(t) = -B^T e^{A^T (t_f-t)} C^T [C W_B C^T]^{-1} C e^{At_f}\mathbf x_0
   \end{array}
\end{equation}
where $W_B= \int_0^{t_f}  e^{A (t_f-t)}B B^T  e^{A^T (t_f-t)}dt$.  Then, the  states  of the target node $S$  could reach  the  origin   at time $t=t_f$, i.e.
\begin{equation}
   \begin{array}{lll}
\mathbf{y}(t_f)=Ce^{At_f}\mathbf x_0-  C W_B C^T    (C  W_B C^T)^{-1} C e^{At_f}\mathbf x_0=\mathbf 0
   \end{array}
\end{equation}

Rearranging the node index of the target node $S$ such that $S=[v_{1}~...~v_{|S|}]^T$. Denote
 $\mathbf  X_1=[x_{1}~...~x_{|S|}]^T$  and
 $\mathbf  X_2=[x_{|S|+1}, ... , x_{N}]^T$ as the state of the target set $S$ and non-target set  $V-S$, respectively,  we have
  \begin{equation}
\begin{array}{lll}
 \left[
  \begin{array}{ccccccccccccc}
\dot{\mathbf  X}_1(t)\\
  \dot{\mathbf  X}_2(t) \\
  \end{array}
\right] =\left[
  \begin{array}{ccccccccccccc}
 A^{11}
  &  A^{12}\\
        A^{21}
  &  A^{22} \\
  \end{array}
\right]   \left[
  \begin{array}{ccccccccccccc}
 \mathbf {X}_1 (t)\\
  \mathbf {X}_2(t) \\
  \end{array}
\right]+  \left[
  \begin{array}{ccccccccccccc}
B_1\\
B_2 \\
  \end{array}
\right] \mathbf u(t) \\
 \end{array}
\label{systemseparated}
\end{equation}
where $A^{11}$ represents the    adjacency matrix of  the target set $S$,  $A^{22}$
the   adjacency matrix  of the  $N-|S|$  non-target nodes in the   set $V-S$. The non-zero entries in $A^{21}$ and
$A^{12}$  represent the  connections  between  $S$ and $V-S$. $B_1$ and $B_2$ are the corresponding input  matrices for $S$  and  $V-S$, respectively.   The  $(A, B, C)$ is target controllable implies that   state variable  $\mathbf{X}_1$  is  structurally
controllable.   \bigskip

 {
 {\it Definition 1.$~~$} {\it
  a) Let   $D(V,E)$   be the diagraph constructed based on $A$.   Denote  $G(A,B)$ as  a digraph  $D(\tilde{V},\tilde{E})$  where the  vertex  set is $\tilde{V}=V \cup  V_B $ and   $\tilde{E}=E \cup E_B$, where     $V_B$ represents  the  $M$  vertices   corresponding to the   $M$  control sources  and  $E_B$ represents   the newly added  edge set  connected to the  control sources  based on  $B$.  A  node $v_i$ in $G(A,B)$ is called inaccessible iff there are no directed paths reaching $x_i$ from  the input vertices $V_B$. A node with a self-loop edge  is  an   accessible node.
    b) The  digraph $G(A, B)$ contains a dilation iff there is a subset $\mathbb{S}\subseteq V$  such that $|T(\mathbb{S})|<|\mathbb{S}|$. Here, the neighborhood set $T(\mathbb{S})$ of a set $S$ is defined as the set of all nodes  $x_j$ where there exists a directed edge from $x_j$ to a  node in $\mathbb{S}$, i.e., $T(\mathbb{S})=\{x_j|(x_j \rightarrow x_i) \in E,x_i \in S\}$, and
   $|T(\mathbb{S})|$  is  the cardinality of set $T(\mathbb{S})$.
  } \bigskip
}

\begin{lem} (Lin's Structural Controllability Theorem \cite{lin1974structural}).
The following  two statements are equivalent ($A$ is the adjacent matrix of the network and $B$ is control matrix of the network):
\begin{enumerate}[\hspace{1em}$a)$]
\item      A linear   system $(A,B)$ is structurally controllable.

\item   The digraph $G(A,B)$ contains no inaccessible nodes or dilation.  $\Box$
\end{enumerate}
  \label{lem1}\end{lem}  \bigskip

Denote  $B=[b_{l_1}, ..., b_{l_i}, ... , b_{l_M}]$ where $l_i$ is a column number index and  $b_{l_i}$   represents    the $l_i$th  column of $B$. We have the following lemma.   \bigskip

\begin{lem}   \cite{blackhall2010structural}   A linear structured system   is structurally controllable  iff there exists a
vertex disjoint union of cacti in the digraph $G(A, B)$   that
covers all the state vertices of  $\mathbf{X}$.  This criteria is satisfied
iff for any  $l_i, l_j\in \{ 1, 2, ... , M\}$  for  $l_i\neq l_j$.

a)  $G({A},b_{l1})$ contains a cactus.

b)  The cacti $G({A}, b_{l_i})$ and $G({A}, b_{l_j})$  are vertex disjoint.

c) The graphs   $G({A}, b_{l_1}), G({A}, b_{l_2}),  ... , G({A}, b_{l_M})$  covers  the state vertices of  $\mathbf{X}$.

If such a vertex disjoint union of cacti exists,  we  call it
a cactus cover. $\Box$
\end{lem}      \bigskip

\begin{figure}
\centering
\includegraphics[width=3.3in]{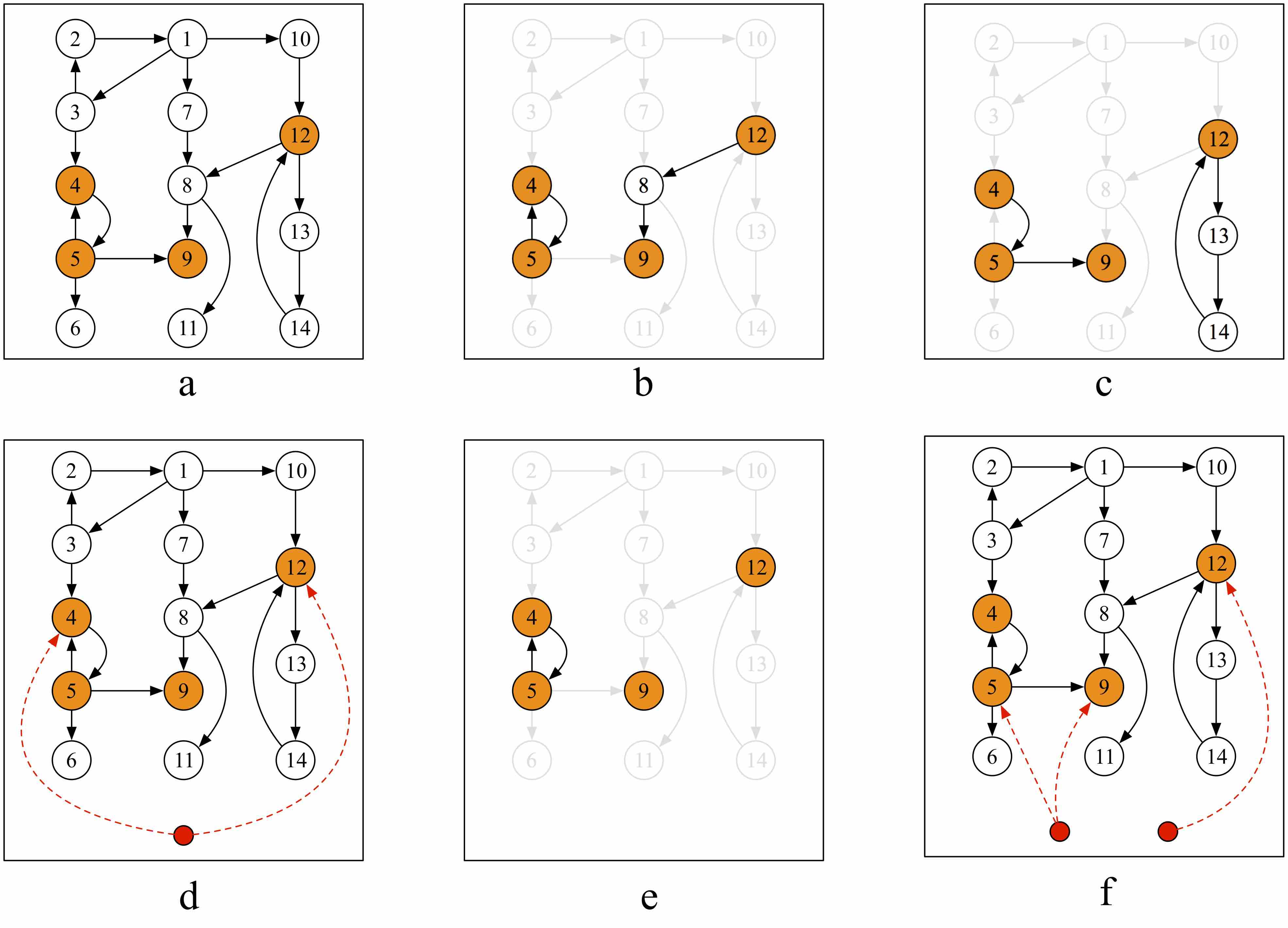}
\caption{ \textbf{An illustration example of the target control problem.}  There are 14 nodes in the network (a).  The subset nodes are colored with orange, where  the target node set is selected as $S=\{v_4, v_5, v_9, v_{12}\}$.  In (b),  a path $v_{12}v_8v_9$ and a circle $v_4v_5v_4$ cover the subset nodes. In (c),  a path $v_{4}v_5v_9$ and a circle $v_{12}v_{13}v_{14}v_{12}$ cover the subset nodes. (d) According to (b) or (c) only one external source colored with red is necessary to control the target subset. (e) However, if we select the control source through employing maximum matching (MM) algorithm, the control paths and circles are depicted in (e). (f) According to (e), the number of external control sources is 2. This  evidently shows that selecting the control sources based on MM algorithm may not achieve the optimal solution.  }  \label{example_bmp}
\end{figure}

 {
\begin{lem}    \cite{blackhall2010structural} The  vertex  set   $S$  corresponding  to   $ \mathbf X_1$  in  equation (\ref{systemseparated})  is  structurally  controllable if $S$ can be  covered by a union of cacti  structure  contained in   the  digraph $G(A, B)$.
\end{lem}   \bigskip


From Lemmas 1-3,  it is known that  when $S=V$  the target controllability problem  is reduced to the  traditional structural  controllability problem,  which can be  well solved either  by   employing  the maximum matching (MM) algorithm.   When   $S\subset V$, the MM algorithm becomes inapplicable and currently there is no algorithm to address this issue as the  target controllability problem  becomes  fundamentally different.
An  example in Figure \ref{example_bmp} is presented to illustrate  that  MM  algorithm does not work for the target control problem.   Therefore, determining  the  minimum  number of  external control sources   for ensuring the target controllability  is  to be investigated in the following Sections.}

\subsection{Converting  the target controllability problem into   the path cover problem in Graph Theory}\label{convertingthetargetcpintopathcover}
For a graph network $D=(V,E)$,
define a set of simple directed paths
$P=\{v_{a_{k,1}} v_{a_{k,2}} ... v_{a_{k,p_k}}| k=1, 2, ...,  |P|\}$
where  $p_k$  is the length of the $k$th   directed  path, $a_{k,i}$ is the index of the $i$th vertex in $V$ along the $k$th directed path
and   $|P|$ is the
total number of paths.  Also,  define a  set of simple directed  circles
$\mathcal{C}=\{v_{b_{k,1}}v_{b_{k,2}} ... v_{b_{k,c_k}}  v_{b_{k,1}}|k=1, 2, ...,  |\mathcal{C}|\}$
 where   $c_k$ is the length of the $k$th   circle,
 $b_{k,i}$ is the index of the vertex on the $k$th circle
  and $|\mathcal{C}|$ is the
total number of circles.
For the  path set
$Q=\{v_{l_{k,1}} v_{l_{k,2}} ... v_{l_{k,q_k}}|k=1, 2, ..., |Q|\}$
 where $q_k$  is the length of the $k$th path, $l_{k,i}$ is the index of the vertex in $V$ along the $k$th path  and    $|Q|$ is the
total number of paths,    define the nodes covered by $Q$ as  $Cover(Q)=\{v_{l_{k,i}} | i=1, 2, ...,  q_k, k=1, 2, ...,  |Q|\}$.

By Lemmas 1-3,  for a given target node set $S$,  the   target controllability problem   can be converted  into the following path cover problem:
\begin{equation}
   \begin{array}{lll}
 argmin_{P}   &|P|^+  \\
s.t.  &  |Cover (P\cup \mathcal{C})|=\sum_{k=1}^{|P|} p_k+ \sum_{k=1}^{|\mathcal{C}|} c_k \\
 & S\subseteq Cover (P\cup \mathcal{C})
 \label{pathcover}
   \end{array}
\end{equation}
where $|P|^+ = max\{|P|, 1 \}$ which is to consist with  the case when
all the  nodes in $S$ exist and only exist in a circle  path.
Therefore what we want to do is to
find one  feasible solution  that $|P|$ is minimal such that
 every node exists and only exists in $P\cup \mathcal{C}$  for at most  once.
Therefore, to  guarantee that all the nodes in  $S$ are target controllable, at least  $|P|$ control sources should be allocated and
 the first node of each directed path should be connected to  a different   control source. Thus,   minimizing the number of control sources is  equivalent to minimizing  $|P|$ in the above path cover problem.


%

\bigskip
\section{Model Transformation to Network Flow Problems}
In the  last  section,    we have  shown  that    the original target controllability problem  is   equivalent to   a path cover problem,  in which the minimum number of external control sources  equals the minimum number of paths.
In this section,   by firstly presenting some  preliminary  knowledge and definitions in Section \ref{definitions},  we shall then propose a  graph transformation method  in Section  \ref{Convertthepathcoverproblem}   to   solve the  path cover problem.   An example is shown in  Figure \ref{testFig}.  In Section \ref{Section3.3},  we  show that the path cover problem can be further transformed  into  the  maximum network flow problem.
In this way,     the target controllability problem can be   solved exactly as a maximum  network  flow  problem.   To verify the validity of the graph transfer, we  prove the equality of the  transformation   in Theorem 3.  Finally in Theorem 4,  it is shown that the   maximum network flow problem   can be solved   within   polynomial time complexity.

\subsection{Preliminary  knowledge and definitions}
\label{definitions}
Some definitions of a network's  maximum flow problem are given as follows
\cite{gross2005graph,gross2004handbook,balakrishnan2012textbook,kleinberg2006algorithm}.  \bigskip

{\it Capacity function:} Given a directed graph $D=(V,E)$, capacity function $c(e)$ is a non-negative function defined in $E$. For an arc $e=(v_i,v_j)$, $c(e)=c_{ij}$ is called the capacity of an arc $e$.  \bigskip

{\it Capacity network:} Given a directed graph $D=(V,E)$ and its capacity function $c(e)$, $D=(V,E,c(e))$ is called the capacity network. $\\$

{\it Flow of the capacity network:} Given a capacity network, flow $f(e)$ is a function defined in $E$. For an arc $e=(v_i,v_j)$, $f(e)=f_{ij}$ is called the flow value on arc $e$, which is bounded by $c(e)$. If $f(e)$ is an integer, we call it an integer flow.  \bigskip

{\it Source, sink and intermediate vertices:} In  a capacity network, the source node  is  denoted as   $v_s$ whose in-degree equals zero. The sink node is  denoted as  $v_t$ whose out-degree equals zero.  All the other nodes  are  called  intermediate vertices.  \bigskip

{\it Capacity constraints :} For every arc $e$ in $E$, its flow $f(e)$ cannot exceed its capacity $c(e)$, i.e., $\forall e=(v_i,v_j)\in E$,
\begin{align}
	0\le f_{ij}\le c_{ij}.
\end{align} $\\$

{\it Conservation constraints:} For every intermediate vertex, the sum of the flows entering it (in-flow) must equal the sum of the flows exiting it (out-flow). Namely $\forall v_i\in V-\{v_s,v_t\}$,
\begin{align}
	\sum_{(v_i,v_j)\in E} f_{ij} - \sum_{(v_j,v_i)\in E} f_{ji} = 0.
\end{align} \bigskip

{\it Feasible flow:} For a capacity network with source and sink, a flow $f=\{f_{ij}\}$ from $v_s$ to $v_t$ is called a feasible flow if flow $f$ satisfies the capacity constraints and conservation constraints simultaneously.  There may be  multiple    source and sink nodes in a network.  The value of a feasible flow $f$ is defined as
\begin{align}
	v(f) = \sum_{(v_s,v_j)\in E} f_{sj} = \sum_{(v_j,v_t)\in E} f_{jt}.
\end{align}
If all $f_{ij}$  are integers, $f$ is called integral feasible flow. \bigskip

{\it Lower bounds and upper bounds:} Given a directed graph $D=(V,E)$,   for   an edge $e\in E$,   the lower bound flow $l(e)$ and upper bound flow $c(e)$  are two non-negative functions defined in $A$ respectively with $l(e)\le c(e)$.  \bigskip

{\it Feasible circulation:} A circulation $f$ is a flow of $D(V,E)$ such that
\begin{align}
	\forall v_i \in V, \sum_{(v_i,v_j)\in E}f_{ij} - \sum_{(v_j,v_i)\in E}f_{ji} = 0.
\end{align}
A feasible circulation is a circulation $f$ of  $D(V,E)$  such that
\begin{align}
	\forall e=(v_i,v_j)\in E, l(e) \le f(e) \le c(e).
\end{align}

\subsection{From the  path cover problem  to the  network  flow problem} \label{Convertthepathcoverproblem}
\begin{figure*}[htbp]
\centering
\includegraphics[width=0.7\textwidth]{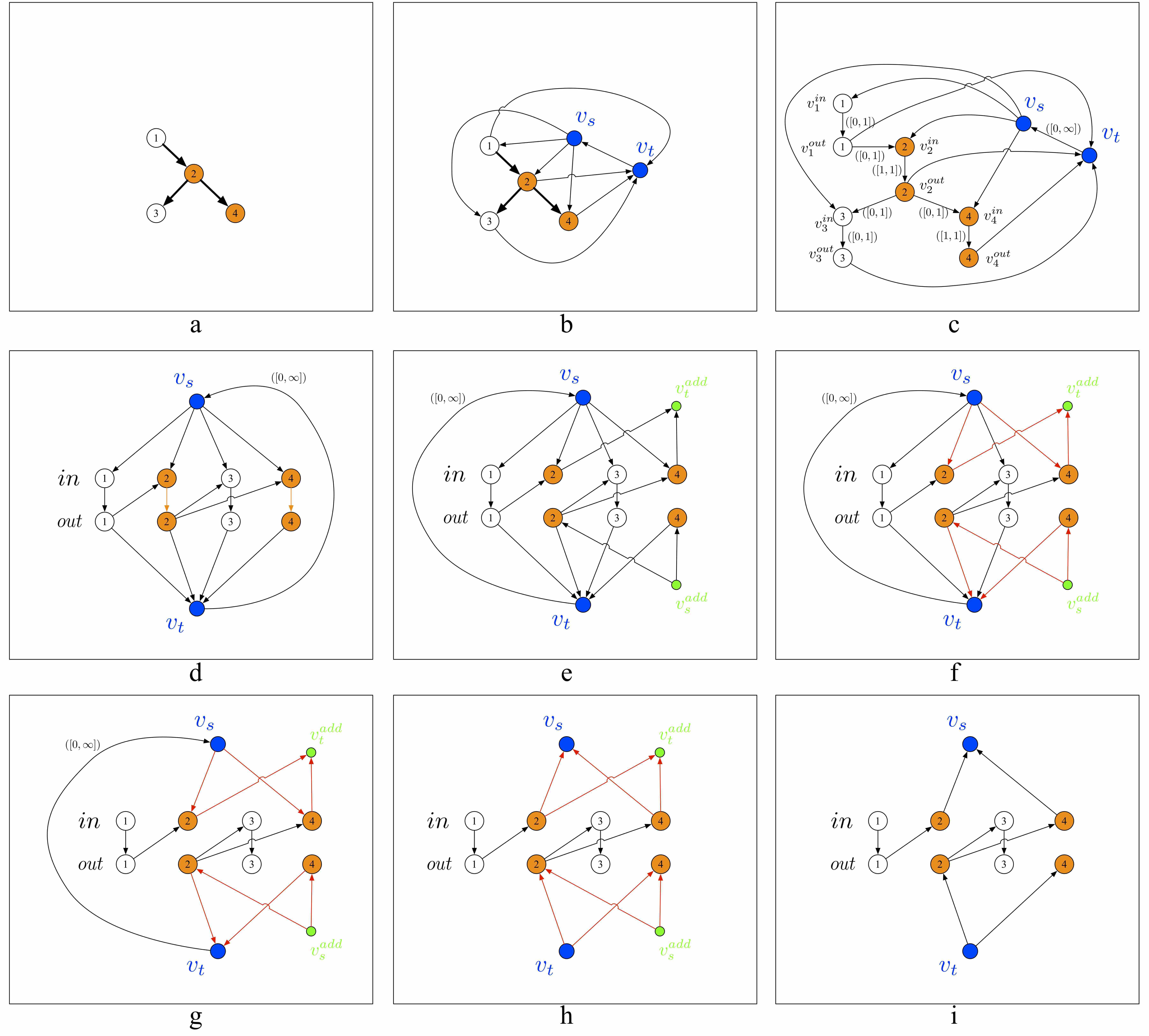}
\caption{\textbf{The procedure of   graph re-construction  to transform  the   problem to  a network  flow problem.} (a) The network we aim to control. It contains 4 nodes and 3 edges with node 2 and node 4 being in the target  set.  (b) By adding a source $v_s$, a sink $v_t$, arcs from $v_s$ to other intermediate vertices and arcs from other intermediate vertices to $v_t$, the network can be converted into a capacity network. (c) After  applying the method of splitting the nodes, the network arc capacities contain lower bounds and upper bounds, which are respectively represented by the left and right numbers in the brackets beside the arcs. (d) By rearranging the position of nodes, the equivalent graph is formed with a source $v_s$, a sink $v_t$, a set of in-nodes and a set of out-nodes. The orange lines from $v_2^{in}$ to $v_2^{out}$ and from $v_4^{in}$ to $v_4^{out}$  are  the arcs with lower bounds $l(a)=1$ and upper bound $c(a)=1$. (e) After applying the graph transfer method of constructing the associate graph, the orange lines are replaced by the arcs from $v_s^{add}$ and to $v_t^{add}$, which transforms  the network with lower bounds and upper bounds  of  the capacity network. (f) The flow of the red lines is 1 and the flow of the black lines is 0. The flows in the graph form a feasible circulation. (g) To find the minimum flow from $v_s$ to $v_t$, the arcs from $v_s$ to $v_1^{in}$ and $v_3^{out}$ and the arcs from $v_1^{out}$ and $v_3^{out}$ to $v_t$ can be removed as their flow is always 0. (h) The solution to the minimum flow problem is equivalent to solving the maximum flow problem from $v_t$ to $v_s$. In the network, the direction of the arcs from $v_s$ to $v_2^{in}$ and $v_4^{in}$ and the arcs from $v_2^{out}$ and $v_4^{out}$ to $v_t$ should be reversed. (i) While finding the maximum flow from $v_t$ to $v_s$, the flow of the arcs from $v_s^{add}$ to $v_2^{out}$ and $v_4^{out}$ and the arcs from $v_2^{in}$ and $v_4^{out}$ to $v_t^{add}$ should be always zero. Thus these arcs can be removed. Finally, the structural control problem of (a) is equivalent to the maximum flow problem from $v_t$ to $v_s$ in (i).}
\label{testFig}
\end{figure*}

 According to the path cover problem described in Section \ref{convertingthetargetcpintopathcover}, we are now endeavoring to find a path set $P$ with the minimal cardinality $|P|$ such that the given target node set $S\subseteq Cover (P\cup \mathcal C) $. To introduce the concept flow into the path cover problem, we artificially add a source $v_s$ and a sink $v_t$ into $D(V,E)$. And for every intermediate vertex $v_i$ in $D(V,E,v_s,v_t)$, there is an arc entering $v_i$ from $v_s$ with $c((v_s,v_i))=1$ and an arc exiting $v_i$ to $v_t$ with $c((v_i,v_t))=1$. For $v_s$ and $v_t$, there is an arc from $v_t$ to $v_s$ with capacity $c((v_t,v_s))=\infty$. For each arc $e\in E$, we set its  capacity as $c(e)=1$. Now we convert a directed network $D(V,E)$ to a capacity network $D(V,E,v_s,v_t,c(e))$. This is illustrated in Figures  \ref{testFig} (a)-(b).

Considering the problem we aim to solve, for every vertex $v_{sub}$ in   $S$,  there should be one and only one path in the path set $P$ covering $v_{sub}$. In the following,
we  will introduce a graph transfer method to ensure  fulfilling this condition.   \bigskip

{
{\it Graph transfer - node splitting:}   We split a node  $v\in V$ into    two  types of  virtual  vertices $v^{in}$  and
 $v^{out}$, respectively.  Therefore, as shown  in Figure \ref{testFig} (c),   the  network  contains three  types  of nodes.   The first type of nodes  are  the source and sink node  $v_s$ and $v_t$.
 The second type  of  nodes  are   $v^{in}_{sub}$  ($v_{sub}\in S$)  and   $v^{in}_{i}$ ($v_{i}\in V-S$) , the original arcs (edges) entering the node $v_{sub}$  or  $v_{i}$ now enter $v_{sub}^{in}$ or $v_{i}^{in}$. Similarly,  the third  type  of nodes correspond to  $v_{sub}^{out}$ and $v_{i}^{out}$, by which the  arcs  exiting node $v_{sub}$ or $v_{i}$   now exit $v_{sub}^{out}$ or $v_{i}^{out}$.
 Besides, for every pair $v_{sub}^{in}$ and $v_{sub}^{out}$, we add an arc $e=(v_{sub}^{in},v_{sub}^{out})$ from $v_{sub}^{in}$ to $v_{sub}^{out}$ with lower bound $l(e) = 1$ and upper bound $c(e)=1$. For other nodes $v_i\in V-S$, the process of splitting is same as that for $v_{sub}$ except that the lower bound is zero, viz. $l(e)=0$. }


Node    splitting converts a  capacity   network into a capacity network with lower bounds and upper bounds $D'=(V',E',v_s,v_t,l(e'),c(e'))$ where $V'=\{v^{in},v^{out}\}$ and $E'=E\cup \{(v^{in},v^{out})\}$. In the  following paragraphs,  we will show how to convert a path cover problem to a  maximum flow problem.   \bigskip

\begin{theorem}
The feasible circulation of the network $D'=(V',E',v_s,v_t,l(e'),c(e'))$ corresponds to a structurally controllable subset $S\subseteq V$ of the linear system $(A,B,C)$.
\label{the2}
\end{theorem} \bigskip

{\it Proof.} The feasible circulation of the capacity network $D$ satisfies  the capacity constraints and the conservation constraints. Specifically,  for the nodes in the subset $S$, we have $$f\left((v^{in}_{sub},v^{out}_{sub})\right)=1.$$  Thus, the flows of the arcs from $v_s$ to the nodes in the subset $S$ are exactly 1. The flows from $v_s$ to $v_t$ are control paths, while the flow from $v_i$ to itself is a circle. Then every node in $S$ is on a certain control path or   circle as $f\left((v^{in}_{sub},v^{out}_{sub})\right)=1$.
According to Lemma \ref{lem1}, the subset of the linear system $G$ is structurally controllable. $\Box$  \bigskip

To show  the existence of the feasible circulation, we introduce another graph transfer method  for  constructing the associate graph, which is shown in Figure \ref{testFig} (e). \bigskip

{\it Constructing the associate graph:} Given a capacity network with lower bounds and upper bounds $$D'=\left(V',E',v_s,v_t,l(e'),c(e')\right),$$ an additional source $v_s^{add}$ and an additional sink $v_t^{add}$ are added into the network. For every node $v_i \in V'$, a new arc $e$ is added from $v_s^{add}$ to $v_i$ with $l(e)=0$ and $c(e)=\sum_{e'=(v_j,v_i)\in E'}l(e')$. And a new arc $e$ is added from $v_i$ to $v_t^{add}$ with $l(e)=0$ and $c(e)=\sum_{e'=(v_i,v_j)\in E'}l(e')$. Meanwhile the original arcs in the network $D'$ decrease their $l(e')$ to zero and $c(e')$ to $c(e')-l(e')$.  The new network, termed  {\it associate graph}  hereafter,   is $D''=(V'',E'',c'(e''))$ with all $l'(e'')\equiv 0$, where
\begin{equation}
\begin{split}
 V''=V'\cup \{v_s^{add},v_t^{add}\},
\end{split}
\end{equation}
\begin{equation}
\begin{split}
 E''=E'\cup \{(v_s^{add},v_i)|v_i\in V'\} \cup \{(v_i,v_t^{add})|v_i\in V'\},
\end{split}
\end{equation}
\begin{equation}
\begin{split}
\forall v_i\in V',~c'(v_s^{add},v_i)& =\sum_{e'=(v_j,v_i)\in E'}l(e'),\\
c'(v_i,v_t^{add})&= \sum_{e'=(v_i,v_j)\in E'}l(e'),
\end{split}
\end{equation}
\begin{equation}
\begin{split}
 \forall e'=(v_i,v_j)\in E',~c'(e') = c(e')-l(e').
\end{split}
\end{equation}
\bigskip

\begin{lem} \cite{gross2005graph}
Iff the value of the maximum flow of the associate graph $D''$ from $v_s^{add}$ to $v_t^{add}$ is equal to the sum of lower bound values of all arcs in the capacity network $D'$, the feasible circulation of $D'$ exists.
\label{associateG}
\end{lem}
\bigskip

\begin{theorem}
The feasible circulation of the network $D'$ converted from the linear system $G(A,B,C)$ always exists.
\label{the3}
\end{theorem} \bigskip

{\it Proof.}  As the source $v_s$ and the sink $v_t$  are  connected to every node in $V$, $v_s$ is connected to every $v^{in}$ and $v_t$ is connected to every $v^{out}$. Thus, there is always a circulation from $v_s^{add}$ to $v_t^{add}$, viz. $v_s^{add}v_{sub}^{out}...v_tv_s...v_{sub}^{in}v_t^{add}$.
Obviously, in the associate graph $D''$, the value of the  maximum flow from $v_s^{add}$ to $v_t^{add}$  equals  the number of the nodes in the subset $S$. In the capacity network $D'$, only the arcs from $v_{sub}^{in}$ to $v_{sub}^{out}$ have the lower bound value $l(e)=1$ while the lower bounds of other acrs are zero. Thus,   the sum of lower bound values of all arcs in $D'$ is also equal to the number of nodes in the subset $S$. According to Lemma \ref{associateG}, the feasible circulation of $D'$ always exists. $\Box$  \bigskip

Based on  the above Lemmas and Theorems,  for a given  target set in $D(V,E)$    that  corresponds to  a  linear system $G(A,B,C)$ in (\ref{systemmodel1}) as illustrated in Figure \ref{testFig} (a),  we can re-construct  a new graph network $D^{(N)}=(V^{(N)},E^{(N)})$,  and   convert it to a capacity network $D^{(N)}=(V^{(N)}, E^{(N)},v_s,v_t,c(e))$.   The procedures  are described    as follows.

\begin{enumerate}[\hspace{1em}a)]
\item  For every node $v\in V$,  split  it  into two nodes    $v^{in}$ and $v^{out}$. Define a node set $V^{(I)}=\{v^{{in}} | v\in V\}$,  $V^{(O)}=\{v^{{out}} | v\in V\}$.

\item Define $V^{(N)}=V^{(I)} \cup V^{(O)} \cup \{v_s, v_t\}$ where
$v_s$  represents  the source and $v_t$    represents    the sink. By setting the upper bound of each edge as $c\left((v_i,v_j)\right)=1$ where $v_i, v_j\in V^{(N)}$ and lower bound as $l((v_{sub}^{in},v_{sub}^{out}))=1$ and $l((v_i,v_j))=0$ where $(v_i,v_j)\notin E-\{(v_{sub}^{in},v_{sub}^{out})\}$, we build a flow network denoted as
 $(D^{(N)}, cap)$ where $cap$ denotes the capacity function.

 \item  {Construct  the  edge sets  $E^{(1)}= \{(v_t, v^{out}) | v\in S\}$ , $E^{(2)}= \{(v^{(in)}, s) |    v\in S\}$,   $E^{(3)}= \{(v^{in}, v^{out}) |    v  \not\in S\}$,  $E^{(4)}= \{(v^{out}_i, v^{in}_j) |   (v_i,v_j)\in E\}$  and $E^{(N)}= E^{(1)} \cup  E^{(2)}   \cup  E^{(3)} \cup  E^{(4)}$.}

     \item   For  each  edge  $(v_i,v_j)\in E^{(N)}$,   set  its upper capacity as $c((v_i,v_j))=1$.
\end{enumerate}
Finally,  by  defining  a  residual  network as $(D^{(N)}, cap, flow)$  where  $flow(e)$ is the flow of edge $e$  such that $l(e)\leq flow(e)\leq c(e)$,    we can    convert the path cover problem into a  maximum  network flow problem in  $(D^{(N)},cap,v_s,v_t)$ as shown in Figure \ref{testFig} (i)  from  $v_s$ to $v_t$.  \bigskip

An example for the  graph re-construction is illustrated in Fig.\ref{testFig} (a)-Fig.\ref{testFig} (i).  For a given  linear system $(A,B,C)$,   the  adjacent matrix $A$ gives the topology information of the capacity network. An arc $e=(v_i,v_j)$ is in the arc set $E$ if $A(i,j) = 1$. The source $v_s$ and the sink $v_t$ can be treated as the same node as they both represent the set of external drivers defined by the control matrix $B$, viz. there is an arc from $v_t$ to $v_s$ with capacity of infinity $c\left((v_t,v_s)\right)=\infty$. As they are external drivers, there  are arcs  from $v_s$ to every vertex in the network and arcs from every vertex to $v_t$.
 As capacity function $c(e)=1,\forall e\in E$, the sum of out-flow of $v_s$ or the sum of in-flow of $v_t$ is equal to the number of external drivers.

  The detailed process of re-constructing the edge sets in the procedure $c)$   is shown in Figs. \ref{testFig} (c)-(i). After the procedure $b)$, we rearrange the position of the nodes as shown in Figs.\ref{testFig} (c)-(d). For this capacity network with upper bounds and lower bounds, we have to re-construct it by means of constructing the associate graph as described before. An additional source $v_s^{add}$ and an additional sink $v_t^{add}$ are added into the network to transform the capacity network with upper and lower bounds to the associate network with its all lower bounds $l(e)=0$ as illustrated in Figure \ref{testFig} (e). Based on the Theorem \ref{the2},
to find the structurally controllable scheme with the minimal external sources,  we start from the feasible circulation, which is shown in Figure \ref{testFig} (f).  While ensuring the feasibility of the circulation in $D'$, the control scheme with the minimal controllers can be achieved as long as the value of flow from $v_s$ to $v_t$ decreases to the minimum. Thus for the vertex $v_i  \in V'-S$, the flow of the arcs from $v_s$ to $v_i^{in}$ or from $v_i^{out}$ to $v_t$ is zero. And the arcs can be removed, which is shown in Figure \ref{testFig} (g). The minimum flow problem from $v_s$ to $v_t$ can be converted to the maximum flow problem from $v_t$ to $v_s$, which is shown in Figure \ref{testFig} (h). In order to find the maximum flow from $v_t$ to $v_s$, the arcs from $v_s^{add}$ or to $v_t^{add}$ can be removed as they cannot contribute to the increase of the flow, which is shown in Figure \ref{testFig} (i).

%
%

\subsection{Maximum-flow based target path-cover  (MFTP) algorithm} \label{Section3.3}
   In this subsection,    we   focus on discussing   how to locate the   circle path set $\mathcal C$  and   the  simple directed  path set $P$  containing the least number of  simple  paths  $|P|$    to cover $S$.  In the previous subsection,   we  carried  out  graph transformation method   to   address the  path cover problem through  solving the  maximum network flow problem $(D^{(N)},cap,v_s,v_t)$ from  $v_s$ to $v_t$.
In the following,  an  algorithm  named  ``{\it maximum-flow based target path-cover''} (MFTP) is first proposed  regarding how to  obtain  the   maximum flow  of $(D^{(N)}, cap, v_s,v_t)$   from $v_t$ to $v_s$  in  Figure \ref{testFig} (i).
In the next subsection, it  will   be shown  in Theorems 3-4 that  the solution of the original target controllability  problem is   equivalent  to  finding  the maximum flow in $(D^{(N)}, cap, v_s,v_t)$ which  can be solved by   introducing the  Dinic algorithm with a  polynomial complexity.
The  MFTP algorithm is presented as follows.
\begin{enumerate}[\hspace{1em}step $1$)]
\item For a given  graph  network $D(V,E)$,  build a new graph network  $D^{(N)}=(V^{(N)}, E^{(N)})$ according to  procedure (a)-(d) described in Section  \ref{Convertthepathcoverproblem}.

\item  Define      $E^{(F)}=\{(v_i,v_j) |  (v^{out}_i, v^{in}_j)\in E^{(4)}, flow((v^{out}_i, v^{in}_j))=1\}$.
    Obtain  the edge set  $E^{(F)}$   by applying  the  Dinic algorithm  \cite{dinits1970algorithms} \cite{dinitz2006dinitz}  to  the  maximum flow problem  in $(D^{(N)}, cap, v_s,v_t)$.   Let  directed path set $P\leftarrow \emptyset$, circle path set $\mathcal{C}\leftarrow \emptyset$,  do the following steps:

\item  For each node $v_i\in S$, if there does not exist a node $v_j \in V$ such that $(v_i,v_j) \in E^{(F)}$ or $(v_j,v_i) \in E^{(F)}$,  then  update
$P\leftarrow P \cup \{(v_i)\}$, i.e., add all the simple paths that contain  only a single node  $v_i$ to $P$.

\item  Find one node $v_1$ such that there  exist  no  node $v_0$ satisfying $(v_0,v_1) \in E^{(F)}$. Continue  the process to find nodes $v_2,...,v_p$ which form a unique sequence $v_1v_2...v_p$ such that $(v_1, v_2), (v_2, v_3), ... , (v_{p-1}, v_p)\in E^{(F)}$ until there does not exist $v_{p+1}$ satisfying that $(v_{p}, v_{p+1})\in E^{(F)}$. Then,  add the path $v_1 v_2 ... v_p$ to $P$, i.e., update   $P\leftarrow P\cup \{v_1 v_2 ... v_p\}$ and delete all the edges $(v_1, v_2), (v_2, v_3), ... , (v_{p-1}, v_p)\in E^{(F)}$.

\item Repeat Step  4 until no more $v_1$  can be found.

\item If there exists any edge in $E^{(F)}$, then for  an arbitrary edge $(v_1, v_2)\in E^{(F)}$,  find a unique sequence $v_3, v_4, ..., v_c$ such that
  $(v_2, v_3), ... , (v_{c-1}, v_c), (v_{c}, v_1)\in E^{(F)}$. As will be proved  in Theorem  3   such node  $v_1$  always exists.
  Then,  add the path $v_1 v_2 ... v_c v_1$ to $\mathcal C$ and update   $\mathcal{C}\leftarrow \mathcal{C}\cup  \{v_1 v_2 ... v_c v_1\} $,  and delete all the edges $(v_1, v_2), (v_2, v_3), ... , (v_{c}, v_1)\in E^{(F)}$.

  \item Repeat the process in Step  6  until $E^{(F)}$ becomes an empty set, and all the simple paths and circles in  $P$ and $\mathcal{C}$ are  finally obtained.
\end{enumerate}

\subsection{Equivalence of the path cover problem and  the maximum network flow problem} \label{Section3.4}
In this  subsection,  we  will  prove  that  the solution of the original target controllability  problem is   equivalent  to  finding  the maximum flow in $(D^{(N)}, cap, v_s,v_t)$ which  can be solved by   introducing the  Dinic algorithm with  a polynomial complexity,  and  the number of the minimum external sources    equals the  minimum number of   simple paths in $P$ given by  $min\{|P|\}= |S|-maxflow(D^{(N)}, cap, s,t)$. \bigskip

\begin{theorem} The target controllability problem  in  the original  graph  network $D(V, E)$   in (\ref{pathcover}) is  equivalent to    the converted  maximum flow problem   $maxflow(D^{(N)}, cap, v_s,v_t)$.
\end{theorem} \bigskip

{\it Proof.} We need to prove the following  four  statements:

1.  Any feasible flow in the  flow network $(D^{(N)}, cap)$ corresponds to a feasible solution of the original target controllability problem in  $D(V, E)$. This is to prove that there exists an injective mapping   from  a feasible residual  network  $(D^{(N)}, cap, flow)$ with integral flow to  the set    $P\cup \mathcal{C}$   such that  $S \subseteq  P\cup \mathcal{C}$  and each element of   $S$  exists and only exists once in  $P\cup \mathcal{C}$. Here, the residual network refers to the network $(D^{(N)},cap)$ with its flow reaching maximum.

As we can build $P$ and $\mathcal{C}$ based on the obtained  $E^{(F)}$ from the above  described    algorithm,  we only need to prove that its solution is a  feasible  solution. Note  that  here we are not discussing on how to  build $P$ and $\mathcal{C}$  when a maximum flow has been  achieved. Instead, we are proving that any  $$E^{(F)}=\{ (v_i,v_j) | (v_i^{out}, v_j^{(in)})\in E^{(4)}, flow((v_i^{out}, v_j^{in}))=1\}$$ corresponding to  a feasible   flow
 gives a  feasible   solution  of  (\ref{pathcover})   in the residual  network  $(D^{(N)}, cap, flow)$ with only integral flow.

  To this end,   firstly, we  prove that

  \begin{enumerate}[\hspace{1em}$a$)]
\item  $\forall v_i\in V:$ there exists at most one $v_j\in V$ such that $(v_i,v_j)\in E^{(F)}$, or there exists at most one $v_j\in V$ such that $(v_j,v_i)\in E^{(F)}$.  This  is   to prove that there exists at most one node $v_j\in V$  such that  $(v_i^{out},v_j^{in})\in E^{(F)}$ and  $flow((v_i^{out},v_j^{in}))=1$.

  Consider  the input edges of a node $v_i^{out}$:  if  $v_i\in S$, then there exists one and only one edge $(v_t, v_i^{out})\in E^{(1)}\subseteq E^{(N)}$; if $ v_i \not  \in  S$, then there exists one and only one edge $(v_i^{in}, v_i^{out}) \in E^{(3)}\subseteq E^{(N)}$. By  combining  both cases, there exists only one edge with  capacity  1 pointing to  $v^{out}_i$. As the flow in the network is always an integer,  there  exists at most one $v_j\in V$  such that  $(v_i^{out},v_j^{in})\in E^{(4)}$ and  $flow((v_i^{out},v_j^{in}))=1$. Similarly, considering that $v_i^{in}$ has at most one output edge belonging to either $E^{(2)}$ or $E^{(3)}$ depending on whether $v_i$ belongs to $S$ or not,  then there exists at most one $v_j\in V$ such that $(v_j,v_i)\in E^{(F)}$.

\item  If  Step 3 of MFTP algorithm can be  processed,  a simple path can always be located in which all nodes are covered  only once.

 In the case that we cannot find a simple path or Step 3 goes into an  infinite loop,  then there must exist a circle path
 and the algorithm goes into
  an infinite loop. In this case, on the path $v_1, v_2, ...., v_p, ...$, there must exist repetitious nodes. Without loss of generality, we assume that the first repetitious node pair is $v_i$ and $v_j$ with ($i<j$), i.e. $v_i$ and $v_j$ are the same node appearing in different places on the path. If $v_i=v_1$, then
  $(v_{j-1}, v_j)=(v_{j-1}, v_i)=(v_{j-1}, v_1)\in E^{(F)}$, which  contradicts     the fact that $v_1$ does not have any input; if $v_i\neq v_1$, then
  $(v_{j-1}, v_j)=(v_{j-1}, v_i)=(v_{i-1}, v_i)\in E^{(F)}$. But it is known that $v_{i-1} \neq v_{j-1}$ as $v_i$ and $v_j$ are the first repetitious node pair, which manifests that $v_i$ has two input edges from $v_{i-1}$ and $v_{j-1}$ respectively. This   contradicts     the fact that the input degree of each node cannot be greater than $1$.  Thus we draw the conclusion that  a simple path can always be located  as long as  Step 3 can be  processed.

 \item   If the Step  4 of MFTP algorithm has been  accomplished, then a simple circle can always  be located by implementing Step 5 .

 At this stage, we cannot find any node that  has  an  output edge but   does not have any  input edge;  otherwise, the proposed algorithm  goes back to Step 3. Note that  the input degree of all nodes is not smaller than 1 while the input degree cannot be greater than 1. Therefore, the remaining nodes  have one  and only one  input edge. Then, by starting from an arbitrary node, we can always come back to this node  and a circle can be  uniquely  found, and the remaining  edges exist in one and only  one circle.

\item   All the nodes exist and only exist once in $P\cup \mathcal{C}$ and $S\subseteq cover(P\cup \mathcal{C})$.

   This is obvious since if a node has any input or output edge existing in $E^{(F)}$, then it must have been deleted in Step 3 or Step 5; otherwise, it  would contradict the fact that all input and output degrees  are  not greater  than $1$. If one node $v\in S$ but it does not have an edge in $E^{(F)}$, then it will be added into $P$ in Step 2.
\end{enumerate}

2.  The flow from $v_t$ to $v_s$ of  the residual  network   $(D^{(N)}, cap, flow)$   equals     $|S|-|P|$.

Based on the conservation constraints in Section 3.1, when the total flow of the  residual  network   $(D^{(N)}, cap, flow)$   equals   $f$, we could
obtain $f$ paths on the flow network:
   \begin{equation}
    \begin{array}{lll}
 v_t v_{p_{1,1}}   v_{p_{1,2}} ...   v_{p_{1,d_1}} v_s, \\
  v_t v_{p_{2,1}}   v_{p_{2,2}} ...   v_{p_{2,d_2}} v_s, \\
  ~~~~~~~~~~~~~~~~~\vdots \\
    v_t v_{p_{f,1}}   v_{p_{f,2}} ...   v_{p_{f,d_f}} v_s.
    \end{array}
    \end{equation}
where all these paths start from $v_t$ and end at $v_s$, and the flow of  an edge  $(v_i,v_j)\in E^{(N)}$  equals   the occurrence number of $(v_i,v_j)$ in all paths.  Thus,  for the $i$-th path, $(v_t, v_{p_{i,1}})\in E^{(1)}, v_{p_{i,1}} \in V^{(O)}$ as $v_t$ only
points to $V^{(O)}$; and  $(v_{p_{i,1}}, v_s)\in E^{(2)}, v_{p_{d_i,1}} \in V^{(I)}$.

Consider that $\forall v^{out}\in V^{(O)}$, its  output edges  all belong to $E^{(4)}$.  Therefore,  as long as $v_{p_{i,j}}\in  V^{(O)}$,
 $(v_{p_{i,j}}, v_{p_{i,j+1}})\in E^{(4)}$, and $v_{p_{i,j+1}}\in V^{(I)}$.
 Also, as there is no output edge of $v_s$,  $v_{p_{i,j}} \neq v_s$.  For $\forall v^{in} \in V^{(I)}$,  the edges starting  from $v^{in}$  belong to either $E^{(2)}$ or
 $E^{(3)}$.  As all the edges in $E^{(2)}$ point to $v_s$,  if  $v_{p_{i,j}} \in V^{(I)}$ and $j< d_i$, then
  $(v_{p_{i,j}}, v_{p_{i,j+1}})\in E^{(3)}$ and  $v_{p_{i,j+1}}\in V^{(O)}$.
   Therefore, for  each path $i$, $d_i$ is an even  number.   And  $\forall   1\leq j\leq \frac{d_i}{2}$, $v_{p_i, 2j-1}\in V^{(O)}$,
  $v_{p_i, 2j}\in V^{(I)}$ and   $(v_{p_i, 2j-1}, v_{p_i, 2j})\in E^{(4)}$;
  $\forall   1\leq j\leq \frac{d_i}{2}-1$,  $(v_{p_i, 2j}, v_{p_i, 2j+1})\in E^{(3)}$. Thus, the $f$ paths can be rewritten as
   \begin{equation}
    \begin{array}{lll}
 v_t v^{out}_{p^{'}_{1,1}}   v^{in}_{p^{'}_{1,2}}   v^{out}_{p^{'}_{1,2}} v^{in}_{p^{'}_{1,3}}  ... v^{in}_{p^{'}_{1,\frac{d_1}{2}+1}} v_s, \\
 v_t v^{out}_{p^{'}_{2,1}}  v^{in}_{p^{'}_{2,2}}   v^{out}_{p^{'}_{2,2}}v^{in}_{p^{'}_{2,3}} ...  v^{in}_{p^{'}_{1,\frac{d_2}{2}+1}}v_s, \\
  ~~~~~~~~~~~~~~~~~~~~~~~~~~~\vdots \\
 v_tv^{out}_{p^{'}_{f,1}} v^{in}_{p^{'}_{f,2}} v^{out}_{p^{'}_{f,2}} v^{in}_{p^{'}_{f,3}} ...   v^{in}_{p^{'}_{1,\frac{d_f}{2}+1}}v_s.
    \end{array}
    \end{equation}
  In addition, we have that
      \begin{equation*}
      \begin{split}
&E^{(F)}\\
=&\left\{
  \begin{array}{ccccccccccccc}
  (v_{p^{'}_{1,1}}, v_{p^{'}_{1,2}}), &(v_{p^{'}_{1,2}}, v_{p^{'}_{1,3}}), &...,&(v_{p^{'}_{1,\frac{d_1}{2}}}, v_{p^{'}_{1,\frac{d_1}{2}+1}})\\
  (v_{p^{'}_{2,1}}, v_{p^{'}_{2,2}}), &(v_{p^{'}_{2,2}}, v_{p^{'}_{2,3}}), &...,&(v_{p^{'}_{2,\frac{d_2}{2}}}, v_{p^{'}_{2,\frac{d_2}{2}+1}})\\
 & & \vdots \\
   (v_{p^{'}_{f,1}}, v_{p^{'}_{f,2}}),    &(v_{p^{'}_{f,2}}, v_{p^{'}_{f,3}}), &..., &(v_{p^{'}_{f,\frac{d_f}{2}}},v_{p^{'}_{f,\frac{d_f}{2}+1}})
  \end{array}
\right\}
\end{split}
    \end{equation*}
Since $E^{(1)}$ and  $E^{(2)}$ are built  only based on the nodes in the subset $S$, and $E^{(3)}$ is only  based on $V-S$,
for each path $i$, we conclude that   $v_{p^{'}_{i,1}}, v_{p^{'}_{i,\frac{d_i}{2}+1}}\in S$ and
$v_{p^{'}_{i,2}}, v_{p^{'}_{i,3}}, ... , v_{p^{'}_{i,\frac{d_i}{2}}}\in V-S$.  As we know that the capacity of edges is 1, all $v_{p^{'}_{i,1}}$ are all different, and all
$v_{p^{'}_{i,\frac{d_i}{2}+1}}$  are   different.

Based on the
inductive method, now   we prove the  following conclusion:   when  the edge set $E^{(F)}$ only contains  the elements of the  first $f$ rows (specifically  the first $d_1+d_2+...+d_f$ elements), we have
$f=|S|-|P|$.

In the first step, we aim to prove by inductive method that by applying the proposed algorithm every time when we add a row into $E^{(F)}$, e.g. the $i$th row is added, we can always find two paths in $P$, one ends at $v_{p^{'}_{i,1}}$ denoted  as $v_{-l_i}v_{-l_i+1}...v_{p^{'}_{i,1}}$ and the other starts from $v_{p^{'}_{i,\frac{d_i}{2}+1}}$ denoted as $v_{p^{'}_{i,\frac{d_i}{2}+1}}...v_{r_i-1}v_{r_i}$.

Firstly, according to the above conclusion that $v_{p^{'}_{i,1}}, v_{p^{'}_{i,\frac{d_i}{2}+1}}\in S$,  when $f=0$, i.e. $E^{(F)}=\emptyset$, $P=\{v| v\in S\}$, we can find two paths in $P$, one ends at $v_{p^{'}_{1,1}}$ denoted as $v_{-l_1}v_{-l_1+1}...v_{p^{'}_{1,1}}$ (actually this path is $v_{p^{'}_{1,1}}$) and the other starts from $v_{p^{'}_{1,\frac{d_1}{2}+1}}$ denoted as $v_{p^{'}_{1,\frac{d_1}{2}+1}}...v_{r_1-1}v_{r_1}$ (actually this path is $v_{p^{'}_{1,\frac{d_1}{2}+1}}$).

Secondly, suppose that after the $k$th row was added into $E^{(F)}$, we can find those two paths as mentioned above. In the following, we are going to prove that if the $(k+1)$th row was added to $E^{(F)}$, we can still find those two paths.

When we add the $k$th row $$\{(v_{p^{'}_{k,1}}, v_{p^{'}_{k,2}}),    (v_{p^{'}_{k,2}}, v_{p^{'}_{k,3}}), ..., (v_{p^{'}_{k,\frac{d_k}{2}}}, v_{p^{'}_{k,\frac{d_k}{2}+1}})\}$$ into $E^{(F)}$, we can find the two paths, one ends at $v_{p^{'}_{k,1}}$ denoted as $v_{-l_k}v_{-l_k+1}...v_{p^{'}_{k,1}}$  and the other starts from $v_{p^{'}_{k,\frac{d_k}{2}+1}}$ denoted as $v_{p^{'}_{k,\frac{d_k}{2}+1}}...v_{r_k-1}v_{r_k}$.

If these two paths are the same, then $v_{-l_k} = v_{p^{'}_{k,\frac{d_k}{2}+1}}$ and $v_{p^{'}_{k,1}} = v_{r_k}$. The proposed algorithm will delete this path from $P$ and add a circle $$v_{p^{'}_{k,1}}v_{p^{'}_{k,2}}...v_{p^{'}_{k,\frac{d_k}{2}}}v_{p^{'}_{k,\frac{d_k}{2}+1}}v_{-l_k+1}...v_{r_k-1}v_{p^{'}_{k,1}}$$ to $\mathcal C$. If these two paths are not the same, the proposed algorithm will delete these two paths from $P$ and add an updated path $$v_{-l_k}v_{-l_k+1}...v_{p^{'}_{k,1}}v_{p^{'}_{k,2}}...v_{p^{'}_{k,\frac{d_k}{2}}} v_{p^{'}_{k,\frac{d_k}{2}+1}}...v_{r_k-1}v_{r_k}$$ to $P$.
Note that we only updated the two or one path we found while all the other paths in $P$ remain unchanged. If the two paths are the same,  $v_{p^{'}_{k,\frac{d_k}{2}+1}}$ will be deleted from the set of starting vertices of all paths in $P$. If the two paths are not the same, the new path still starts from $v_{-l_k}$, $v_{p^{'}_{k,\frac{d_k}{2}+1}}$ is also deleted from the set of starting vertices of all paths in $P$. This is also valid for the path ending at $v_{p^{'}_{k,1}}$.

In general, we only delete $v_{p^{'}_{k,\frac{d_k}{2}+1}}$ from the set of starting vertices of all paths in $P$ and $v_{p^{'}_{k,1}}$ from the set of ending vertices of all paths in $P$. According to the conclusion that  $v_{p^{'}_{i,1}}, v_{p^{'}_{i,\frac{d_i}{2}+1}}\in S$, all $v_{p^{'}_{i,1}}$ are all different in all paths, and all
$v_{p^{'}_{i,\frac{d_i}{2}+1}}$  are   different, $v_{p^{'}_{k+1,\frac{d_{k+1}}{2}+1}}\in S$ is still in the set of starting vertices of all paths in $P$ and $v_{p^{'}_{k+1,1}}\in S$ is still in the set of ending vertices of all paths in $P$, which implies that after the $(k+1)$th row is added to $E^{(F)}$, the two paths can  also be found.

Now we have proven that those two paths can always be  found  each  time we add a row into $E^{(F)}$.

In the second step, we aim to prove that $f=|S|-|P|$ is valid when we add the first $f$ rows into $E^{(F)}$.

Firstly, when $f=0$, i.e. $E^{(F)}=\emptyset$, we have $|P|=|S|$ as $P=\{v| v\in S\}$.

Secondly, suppose that $k-1=|S|-|P|$ when   $E^{(F)}$ contains $k-1$ rows.  In the following, we are going to prove that, if  one row
$$\{(v_{p^{'}_{k,1}}, v_{p^{'}_{k,2}}),   (v_{p^{'}_{k,2}}, v_{p^{'}_{k,3}}), ..., (v_{p^{'}_{k,\frac{d_k}{2}}},v_{p^{'}_{k,\frac{d_k}{2}+1}}) \}$$
is added to
$E^{(F)}$,  we have  $k=|S|-|P|$, which implies that $|P|$ is reduced by  1 in this case.

To avoid confusion,  let  $j=\frac{d_k}{2}$.  According to the proof in the first step, we can always find two paths in the existing $P$, one ends at $v_{p^{'}_{k,1}}$ denoted as $v_{-l} v_{-l+1} ...  v_{p^{'}_{k,1}}$  and the other starts from $v_{p^{'}_{k,j+1}}$ denoted as
 $v_{p^{'}_{k,j+1}} ... v_{r-1} v_r$.
 And according to the first step, if these two paths are the same path, the proposed algorithm will delete this path from $P$ and add a circle $v_{p^{'}_{k,1}} v_{p^{'}_{k,2}} ... v_{p^{'}_{k,j}}v_{p^{'}_{k,j+1}}v_{-l+1} ... v_{r-1} v_{p^{'}_{k,1}}$ to $\mathcal{C}$ . Then $|P|$ will be reduced by $1$. If these two paths are not the same,
 our  proposed algorithm will  delete these two paths from $P$ and add  an  updated path $$v_{-l} v_{-l+1} ... v_{p^{'}_{k,1}} v_{p^{'}_{k,2}} ... v_{p^{'}_{k,j+1}}...v_{r-1} v_r$$ to $P$.  Then  $|P|$ will be also reduced by $1$.  Thus,  $f=k=|S|-|P|$ is still valid.

 Finally, we obtain that $f=|S|-|P|$.

3. Any feasible  solution of the original problem  corresponds to an integer feasible flow in the  flow network $(D^{(N)}, cap)$, and  $|S|-|P|$ is no larger than the flow. This is to prove that  there exists   an injective mapping   which maps a  path cover $P\cup \mathcal{C}$ in which all nodes in subset $S$ exist and only exist once in a  residual  network   $(D^{(N)}, cap, flow)$   with integer flow, and the network flow is not smaller  than $|S|-|P|$.

Firstly,  for each  simple path in $P$, delete the  first and last  few nodes   that do not belong to $S$ such that the first and the end nodes belong to $S$. In the case that  all the nodes on the path do  not belong to $S$, we delete the whole path. It is seen that such operations do not change the fact that all the nodes in $S$  appear and only appear  once in the cover  $P\cup \mathcal{C}$.

Secondly, we construct the   feasible   flow based on transposing the steps as discussed before: we open all the paths and circles at the nodes belonging to $S$ such that the first and last nodes of all new paths belong to $S$. Then, a feasible   flow can be  constructed   from $v_s$ to $v_t$
for   each  simple path. Finally,  feasible    circulations   can be constructed
based on   those nodes that are   not in $S$.

4. The maximum flow is the optimal solution of original problem.

From the   network flow theory, if   the flow capacities of all  the links are integers, then for a given  integer   flow from $v_s$
to $v_t$, there exists an  integer feasible   flow , and it can be proven that the maximum flow is also an integer.

At the same time, the maximum flow algorithm guarantees   that the maximum flow  has been  obtained from  $v_t$ to $v_s$.

Based on  the proofs of sub problems 1 and 2 above,  we could  obtain   all the paths and circles that cover the subset with the network  flow being $|S|-|P|$.  On the other hand, there does not exist a solution  with a smaller  $|P|$;  otherwise, we should have obtained  a   solution which corresponds to feasible flow being $|S|-|P|$. This   contradicts  with the maximum flow theory.

Based on the four proofs above,  we  prove  the existence and optimality of the proposed  method.  $\Box$  \bigskip

 \begin{theorem}  Finding  the optimal solution of the Maximum Flow Problem in MFTP   by Dinic
  algorithm  has a time complexity of      \(O\left( |V|^{1/2} |E| \right)\).
 \end{theorem}  \bigskip

{\it Proof.} The proof can be seen in  \cite{even1975network}. The time complexity of original
Dinic's is \(O\left( |V|^{2}|E| \right)\) which is
fast enough for most graphs. Furthermore, the Dinic's itself can be
optimized to \(O\left( |V||E|\log |V| \right)\) with a data structure
called dynamic trees.

When the edge capacities are all equal to one, the algorithm has a complexity of  $O(|V|^{2/3} |E|)$, and if  the  vertex  capacities are all equal to one, the algorithm has a complexity of  $O(|V|^{1/2} |E|)$.   Since the graph is transformed from the network with unit capacities for all vertices other than the source and the sink, all edge capacities are equal to $1$ and every vertex $v$ other than $s$ or $t$ either has a single edge emanating from it or has a single edge entering it. According to \cite{even1975network}, this kind of network is of {\it type 2} and the time complexity of Dinic's can be reduced into
 \(O\left( |V|^{1/2} |E| \right)\),
 i.e. we can solve the Problem 2 which is more general in this
 complexity. $\Box$ \bigskip

{
In  \cite{liu2011controllability}, it is known that  we can  solve the structural controllability of  the entire  network   by employing MM algorithm  with complexity  $O( |V|^{1/2}|E|)$.
As stated above, our MFTP algorithm has the same time complexity but can deal with more general cases. Actually, a maximum matching problem itself can be solved by transforming it into a network flow problem. Therefore, MFTP is consistent with the MM algorithm when $S=V$ and can be applied to more complicated cases where $S\subset V$ or there are multiple layers in the network.}

\bigskip

\section{Experimental Results}
\subsection{Illustration of target control in a simple network}
\begin{figure}[h!]
\centering
\includegraphics[width=1.2in]{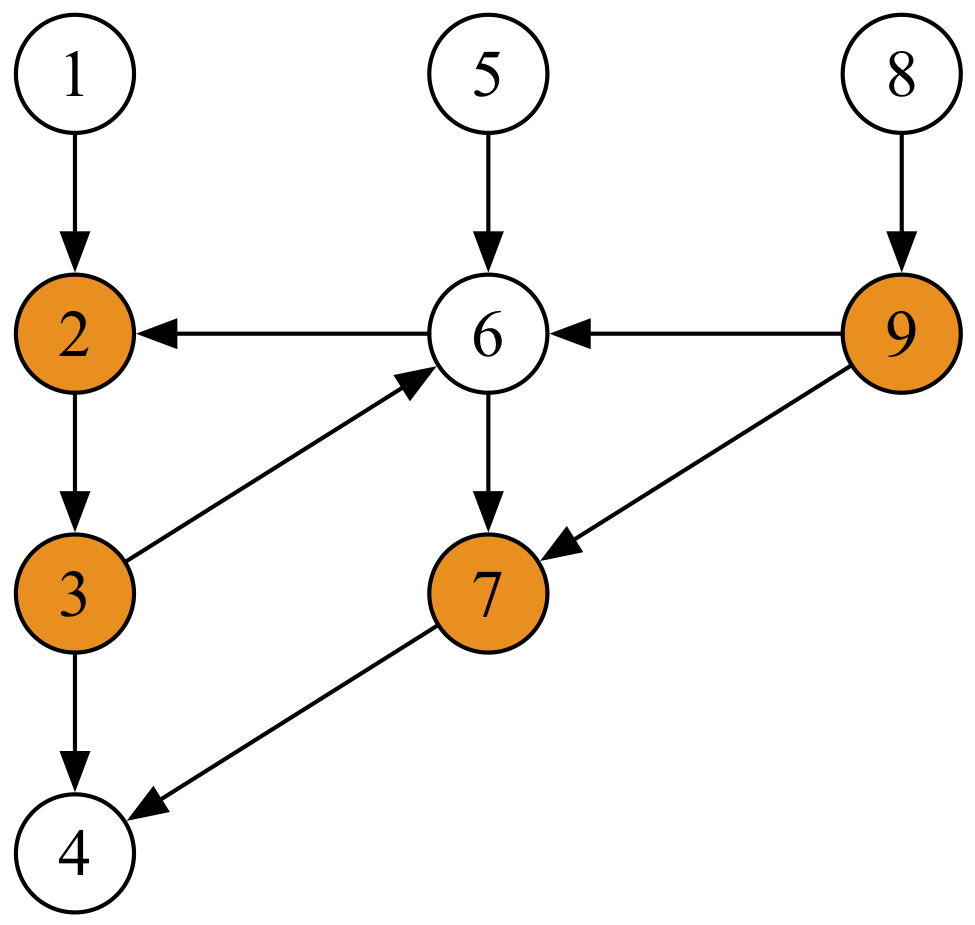}
\caption{ \textbf{A simple example of the target control problem.} There are $9$  nodes in the network with the target set $S$ being  $\{node~2, node~3, node~7, node~9\}$ colored with orange.   }  \label{example_bmp1}
\end{figure}

Consider a simple   example  in Figure  \ref{example_bmp1},  the target node set is selected as $S=\{node~2, node~3, node~7, node~9\}$.  The   matrices  $A$ and $C$ are  given by
\small
\begin{equation*}
\begin{split}
 A=\left(
  \begin{array}{cccccccccccccccccc}
0 &0  &0  &0  &0 &0  &0  &0  &0  \\
1 &0  &0  &0 &0  &1  &0  &0 &0 \\
0 &1  &0  &0 &0  &1  &0  &0 &0 \\
0 &0  &1   &0 &0  &0  &1  &0 &0 \\
0 &0  &0   &0 &0  &0  &0  &0 &0 \\
0 &0  &0   &0 &1  &0  &0  &0 &1 \\
0 &0  &0   &0 &0  &1  &0  &0 &1 \\
0 &0  &0   &0 &0  &0  &0  &0 &0 \\
0 &0  &0   &0 &0  &1  &0  &1 &0 \\
  \end{array}
\right), \\
 C=\left(
  \begin{array}{cccccccccccccccccc}
0 &1 &0  &0  &0 &0  &0  &0  &0  \\
0 &0  &1  &0 &0  &0  &0  &0 &0 \\
0 &0  &0  &0 &0  &0  &1  &0 &0 \\
0 &0  &0   &0 &0  &0  &0  &0 &1 \\
  \end{array}
\right).
\end{split}
\end{equation*} \normalsize

\begin{figure}
\centering
\includegraphics[width=3in]{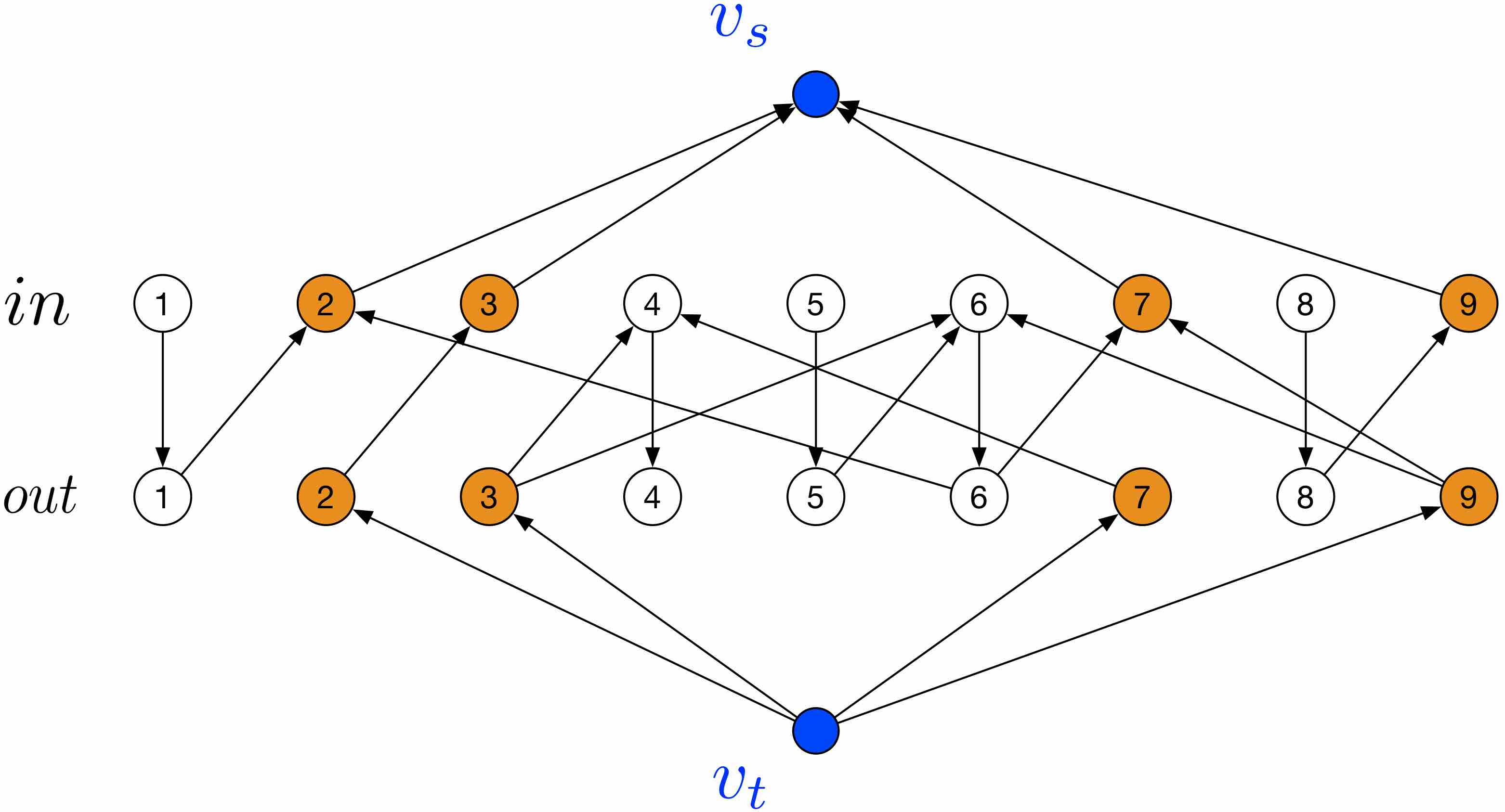}
\caption{ \textbf{ The reconstructed flow  network.} } \label{example_bmp2}
\end{figure}

Recall that the  objective is to  allocate the  minimum number of   external control sources  such that $S$ is controllable.
As shown in Section  4, this problem  can be converted  into  the  maximum flow problem  on the   reconstructed flow  network in Figure \ref{example_bmp2} {similar to Figure \ref{testFig} }.

{Here is a step by step explanation of how this graph is constructed:
\begin{enumerate}
\item For the network in Figure \ref{example_bmp1}, we first use the graph transfer methods \emph{node splitting} to get a graph with node set $V^{(N)}=V\cup V^{(I)}\cup V^{(O)}\cup {v_s,v_t}$ and edge set $E^{(N)}=E\cup \{(v_s,v_i^{in})\}\cup\{(v_i^{out},v_t)\}$.
\item Then use the graph transfer methods \emph{associate graph construct} and reverse the direction of edges in $\{(v_s,v_i^{in})\}\cup\{(v_i^{out},v_t)\}$ to get the graph shown in Figure \ref{example_bmp2}.
\item After this, use Dinic algorithm to get the maximum flow from $v_t$ to $v_s$ which is obviously
\begin{equation*}
f(e)  =
\begin{cases}
1, e \in \left\{\begin{array}{c}
(v_t,v_2^{out}),(v_2^{out},v_3^{in}),(v_3^{in},v_s),\\
(v_s,v_9^{out}),(v_9^{out},v_7^{in}),(v_7^{in},v_s)\\
(v_t,v_3^{out}),(v_3^{out},v_6^{in}),(v_6^{in},v_6^{out}),\\
(v_6^{out},v_2^{in}),(v_2^{in},v_s)
\end{array}\right\}\\
0, others
\end{cases}.
\end{equation*}
\item Thus according to step 2 of MFTP, $E^{(F)}=\{(v_2,v_3),(v_7,v_9),(v_3,v_6),(v_6,v_2)\}$. We set $P\gets \emptyset$ and $\mathcal C\gets \emptyset$.
\item Then we add path $v_9v_7$ into $P$ according to step 4 and step 5 of MFTP, and add $v_2v_3v_6v_2$ into $\mathcal C$ according to step 6 and step 7 of MFTP.
\end{enumerate}
}

\begin{figure}
\centering
\includegraphics[width=2.5in]{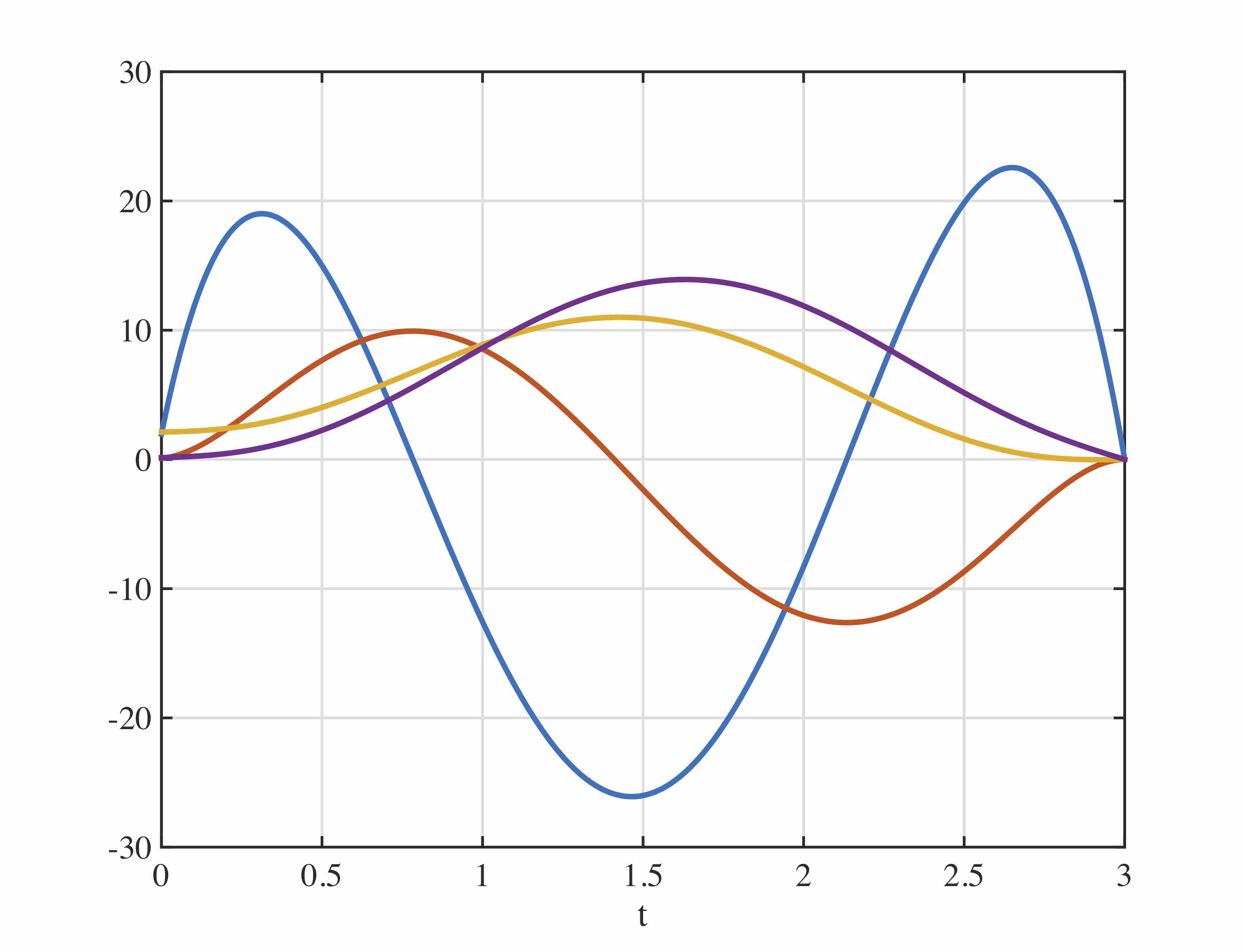}
\caption{ \textbf{}{Illustration of the convergence of the system states.}  The time constant $t_f$ is set as $t_f=3$.   }  \label{example_bmp13}
\end{figure}

Since   target controllability can be guaranteed as long as the nodes in $S$  are covered by a cactus structure when  the matrix $B$ is chosen as $B=[0~ 1~ 0~ 0~ 0~ 0~ 0~ 0~ 1]^T$. { As we  obtain $P=\{node~ 9, node~ 7 \}$, $\mathcal{C}=\{node ~2, node~ 3,  node~ 6\}$, all the nodes covered
by the  cactus $P\cup \mathcal{C}$ ($|P|+|\mathcal{C}|=5$) structure  are  controllable.} As $|P|=1$, there is   only one required  control source, and   $node~ 9$  together with   one node in $\mathcal{C}$   shall  be connected to the external control source, say $node ~2$ for example.  In this case,  the  input    matrix  $B$  is set  as $B=[0~ 1~ 0~ 0~ 0~ 0~ 0~ 0~ 1]^T$, and the input $\mathbf{u}(t)$  can be designed  based on (2).  By doing this,      the    states  of all nodes in $S$  are plotted
in Figure \ref{example_bmp13}. It  is seen that  all the states  approach  the original point at time $t=t_f$. This verifies the
effectiveness of our method.

\subsection{Target control  in  ER, SF and real-life   networks. }
\begin{figure}
\centering
\includegraphics[width=3.3in]{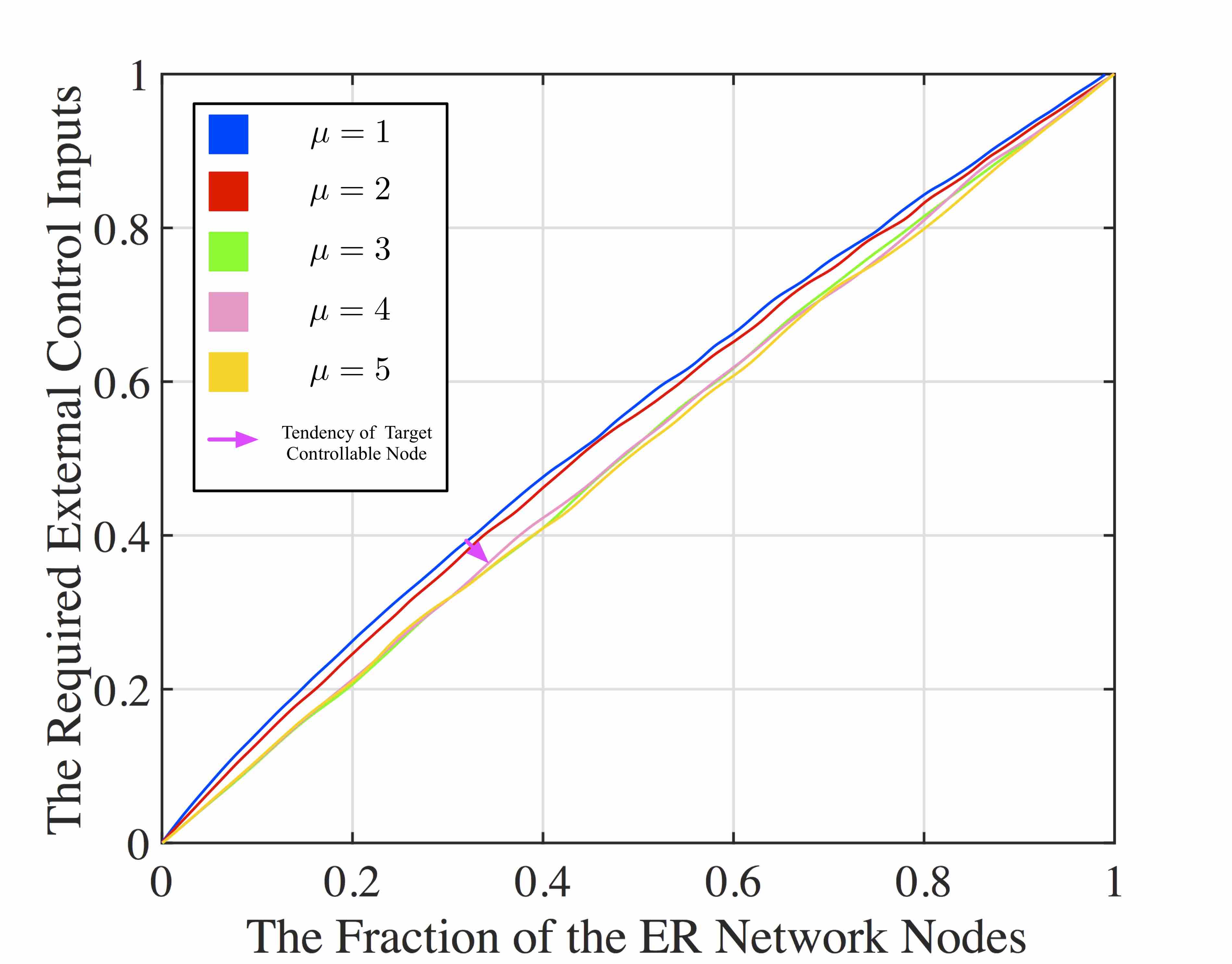}
\caption{ \textbf{Minimum number of required external control sources in ER networks.} {The y-axis is normalized by the total minimum number of driver nodes of the whole network, i.e. the y-axis is the ratio ($\frac{n_D}{N_D}$, where $n_D$ is the minimum number of driver nodes of target subset)}.  }  \label{ERnetworktarget}
\end{figure}

\begin{figure}
\centering
\includegraphics[width=3.3in]{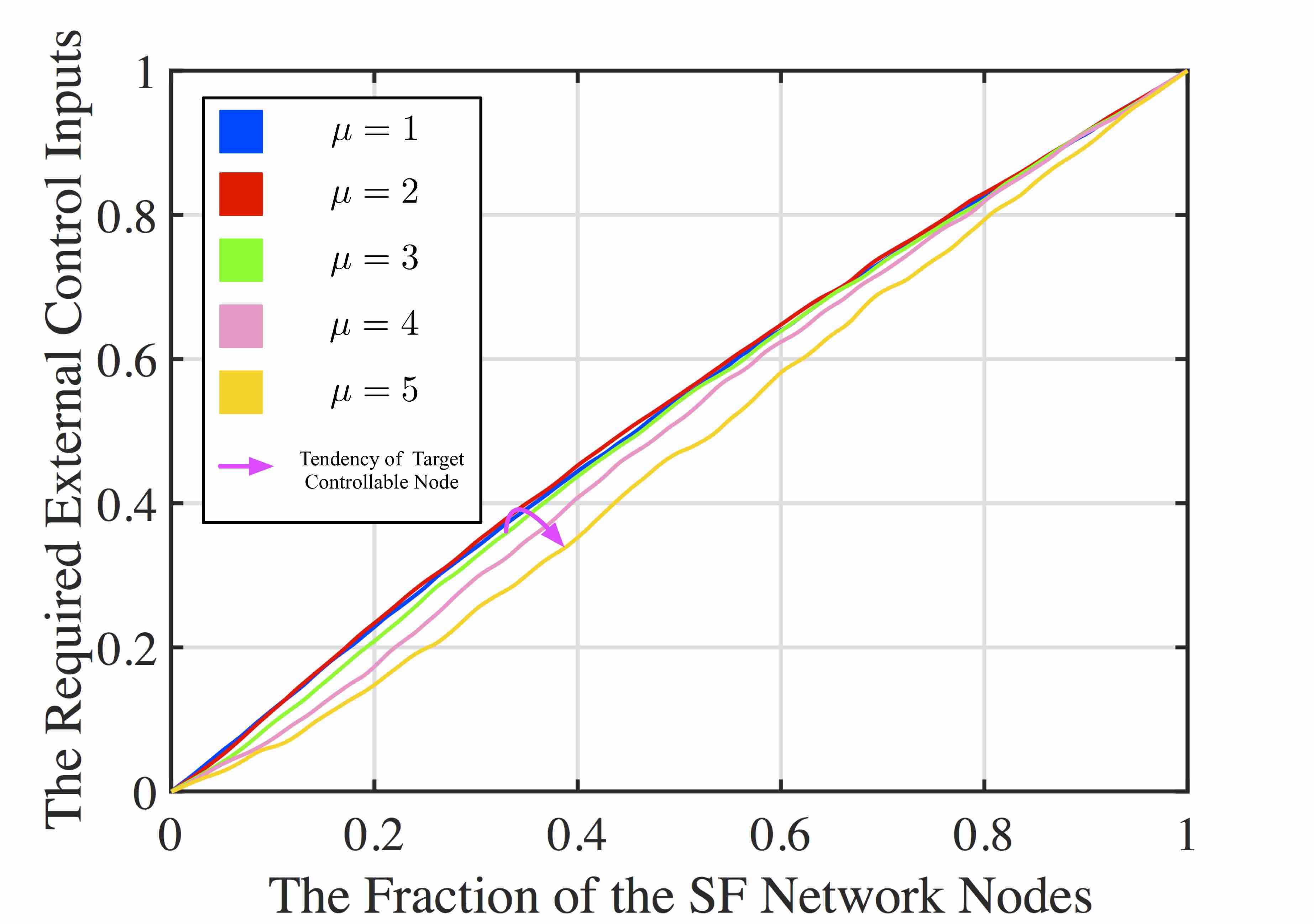}
\caption{ \textbf{Minimum number of required external control sources in   SF  networks.}  {The y-axis is normalized by the total minimum number of driver nodes of the whole network, i.e. the y-axis is the ratio ($\frac{n_D}{N_D}$, where $n_D$ is the minimum number of driver nodes of target subset). } }  \label{SFnetworktarget}
\end{figure}

\begin{figure}
\centering
\includegraphics[width=3.3in]{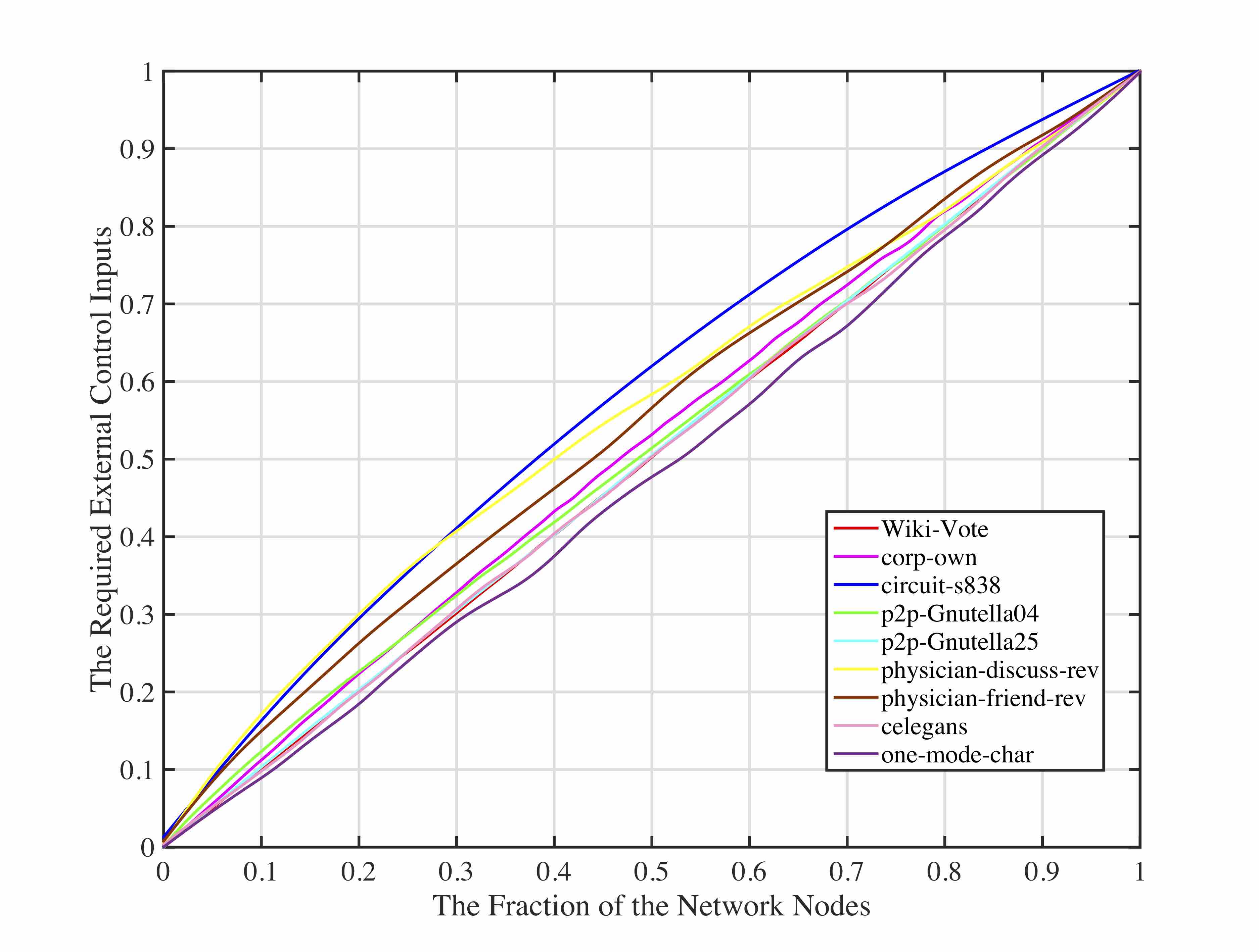}
\caption{ \textbf{Results in real life  networks.}  {The y-axis is normalized by the total minimum number of driver nodes of the whole network, i.e. the y-axis is the ratio ($\frac{n_D}{N_D}$, where $n_D$ is the minimum number of driver nodes of target subset). } }  \label{reallifenetworktarget}
\end{figure}

 In this subsection, we    test    {MFTP\footnote{codes are available on GitHub site: https://github.com/PinkTwoP/MFTP} } in  Erdos-Renyi  (ER) \cite{erdos1960evolution}    networks and Scale-Free (SF)   \cite{ealbert2002statistical} networks  as well as some real-life networks.  The reason of choosing ER and SF networks is because  both of them  preserve  quite common properties  of a vast number of natural and artificial networks.
 Figure \ref{ERnetworktarget} shows the  results in  ER  networks   with $N=1000$ nodes and  $\mu$  varying from $\mu=1$ to $\mu=5$, {where $\mu$ is the mean degree of the network hereafter.}   For a given fraction of network nodes selected as target nodes (denoted as $S$),  the  required minimum number of external control sources for this  case  is  close  to the neutral expected  fraction of external control sources.
That is to say,  to
control an $f$  fraction of target nodes we need  approximately about $f N_D$ external control sources, where $N_D$ is the minimum number of driver nodes using MM \cite{liu2011controllability} algorithm when $S=V$.  As shown in Theorem 4, these   external control sources can be located based on MFTP. We would like to
note that this  conclusion is still valid when  SF
 networks are tested with $\mu=3$, and $\gamma=3$ ($N=1000$) as shown  in Figure \ref{SFnetworktarget}, {where $\gamma$ is the tail index of SF networks.} However, generally SF networks require  more external control sources than ER networks.   Because   typically a SF  network has a
much larger portion of low-degree nodes compared to
an ER network made up of the same volume of nodes
and links. This may lead to significant differences in the
external control sources allocation.  We   have also tested  MFTP  in a few real life networks (Wiki-Vote \cite{leskovec2010signed}, Crop-own \cite{norlen2002eva}, Circuit-s838 \cite{milo2004superfamilies} p2p-Gnutella \cite{ripeanu2002mapping} physican-discuss-rev \cite{leskovec2010predicting}, physican-friend-rev \cite{leskovec2009community}, celegans  \cite{young1993organization}  and one-mode-char \cite{opsahl2009clustering})   as shown in Figure  \ref{reallifenetworktarget}.
It is  observed that    different
network topologies may lead to significant differences  when locating the
target controllable nodes.  Such observations may be important if one want to
understand how the structures of the networks affect the
target control of real-life networks.

\bigskip
\section{Discussions and Conclusion}
In this work, we  have solved   an  open problem   regarding how   to  allocate  the  minimum number
of sources for ensuring the target controllability of  a subset  of nodes  $S$ in   real-life networks in which loops are generally exist.   The target controllability  problem is   converted
to a maximum  flow problem  in graph theory under  specific constraint conditions. We  have  rigorously
 proven    the  validity   of the model transformation.  An  algorithm termed
  ``maximum flow based target path cover'' (MFTP) was  proposed to solve the  transformed problem.
Experimental  examples  demonstrated   the effectiveness of MFTP.

It is  shown  that the solution of the  maximum network  flow problem  provides strictly the  minimum number of  control sources  for
  arbitrary directed networks,   whether the loops exist or not.
By this work,   a link from target structural controllability to network flow problems has been established.
 We anticipate that our work would serve wide applications in  target control of real-life  complex networks,  as well as counter control of various systems which may contribute to enhancing system robustness and resilience.
 As seen in this work, our model considers only LTI systems and we  believe that extending the   results   to  directed  networks with  nonlinear dynamics light up the way of our future research.

.



\ifCLASSOPTIONcaptionsoff
  \newpage
\fi

\bibliographystyle{IEEEtran}
\bibliography{Subnet_20170820_tang}

\begin{IEEEbiography}
[{\includegraphics[width=1in,height=1.25in,clip,keepaspectratio]{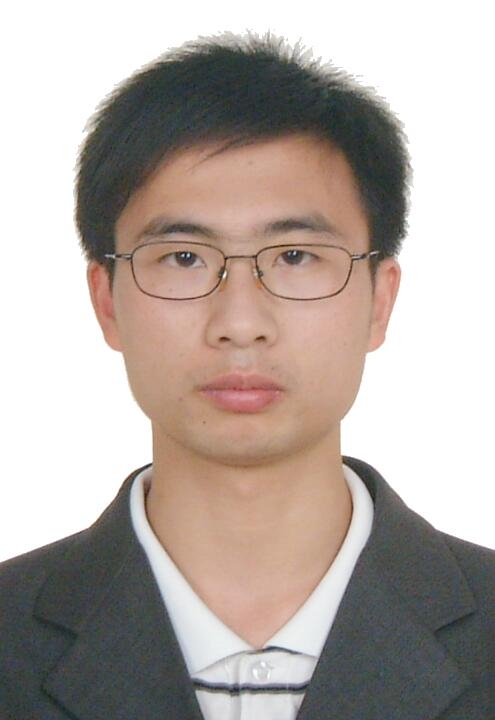}}]
{Guoqi Li} received the B.Eng. degree and M.Eng. degree from Xi'an University of Technology and Xi'an Jiaotong University, P. R. China, in 2004 and 2007, respectively, and Ph.D. degree from Nanyang Technological University, Singapore in 2011.

He was a Scientist with Data Storage Institute and Institute of High Performance Computing, Agency for Science, Technology and Research (A*STAR), Singapore, from September 2011 to March 2014. Since March 2014, he has been an Assistant Professor with the Department of Precision Instrument, Tsinghua University, P. R. China.

Dr. Li has authored or co-authored more than 80 journal and conference papers. His current research interests include brain-inspired computing, complex systems, machine learning and  neuromorphic computing. He has been actively involved in professional services such as serving as an International Technical Program Committee Member and a Track Chair for international conferences. He is an Editorial-Board Member and a Guest Associate Editor for Frontiers in Neuroscience (Neuromorphic Engineering section). He serves as a reviewer for a number of international journals.

\end{IEEEbiography}

\begin{IEEEbiography}
[{\includegraphics[width=1in,height=1.25in,clip,keepaspectratio]{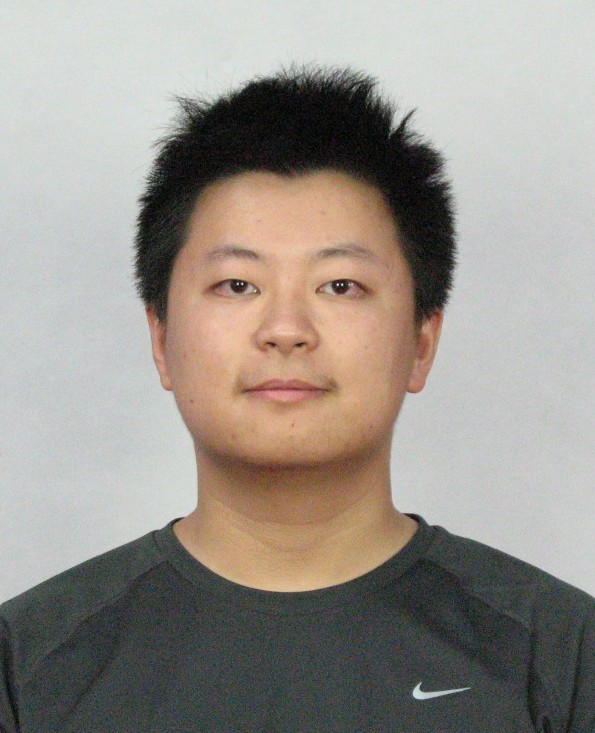}}]
{Xumin Chen} was born in China in 1992. Currently he is a Ph.D. student of Department of Computer Science and Technology in Tsinghua University, P.R.China. His research interests include network representation learning and network mining. He was the main coach of China national team to International Olympiad in Informatics in 2016.
\end{IEEEbiography}

\begin{IEEEbiography}
[{\includegraphics[width=1in,height=1.25in,clip,keepaspectratio]{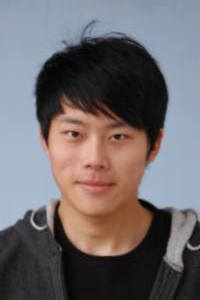}}]
 {Pei Tang}   received the bachelor's  degree from
Tsinghua University, Beijing, China, in 2010, where
he is currently pursuing the Ph.D. degree with
the Center for Brain Inspired Computing Research,
Department of Precise Instrument.
His current research interests include brain
inspired computing, complex systems, reinforcement
learning, and brain inspired chip design.
\end{IEEEbiography}

\begin{IEEEbiography}
 [{\includegraphics[width=1in,height=1.25in,clip,keepaspectratio]{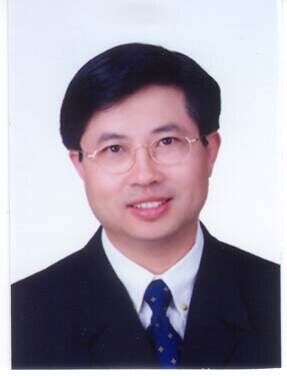}}]
 {Changyun Wen} received B.Eng. degree from Xi'an Jiaotong University, China in 1983 and Ph.D. degree from the University of Newcastle, Australia in 1990. From August 1989 to August 1991, he was a Postdoctoral Fellow at University of Adelaide. Since August  1991, he has been with School of EEE, Nanyang Technological University, where he is currently a Full Professor.  His main research activities are control systems and applications, intelligent power management system, smart grids,  model based online learning and system identification.

He is an Associate Editor of a number of journals including Automatica, IEEE Transactions on Industrial Electronics and IEEE Control Systems Magazine. He is the Executive Editor-in-Chief, Journal of Control and Decision. He served the IEEE Transactions on Automatic Control as an Associate Editor from January 2000 to December 2002. He has been actively involved in organizing international conferences playing the roles of General Chair, General Co-Chair, Technical Program Committee Chair, Program Committee Member, General Advisor, Publicity Chair and so on. He received the IES Prestigious Engineering Achievement Award 2005 from the Institution of Engineers, Singapore (IES) in 2005.

Dr. Wen is a Fellow of IEEE, a member of IEEE Fellow Committee from 2011 to 2013 and a Distinguished Lecturer of IEEE Control Systems Society from February 2010 to February 2013.
\end{IEEEbiography}

\begin{IEEEbiography}
 [{\includegraphics[width=1in,height=1.25in,clip,keepaspectratio]{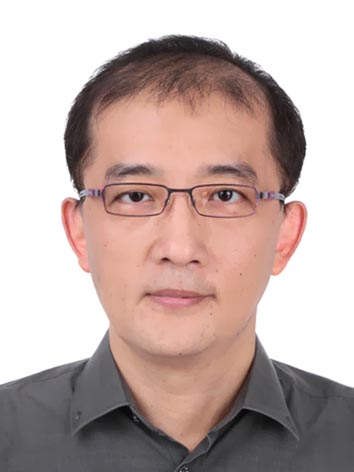}}]
{Gaoxi Xiao}   received the B.S. and M.S. degrees in applied mathematics from Xidian University, Xi'an, China, in 1991 and 1994 respectively. He was an Assistant Lecturer in Xidian University in 1994-1995. In 1998, he received the Ph.D. degree in computing from the Hong Kong Polytechnic University.

He was a Postdoctoral Research Fellow in Polytechnic University, Brooklyn, New York in 1999; and a Visiting Scientist in the University of Texas at Dallas in 1999-2001. He joined the School of Electrical and Electronic Engineering, Nanyang Technological University, Singapore, in 2001, where he is now an Associate Professor. His research interests include complex systems and complex networks, communication networks, smart grids, and system resilience and risk management.

Dr. Xiao serves/served as an Editor or Guest Editor for IEEE Transactions on Network Science and Engineering, PLOS ONE and Advances in Complex Systems etc., and a TPC member for numerous conferences including IEEE ICC and IEEE GLOBECOM etc.
\end{IEEEbiography}

\begin{IEEEbiography}
 [{\includegraphics[width=1in,height=1.25in,clip,keepaspectratio]{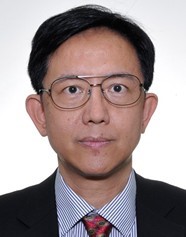}}]
 { Luping Shi} received the B. S. degree and M. S. degree in Physics from Shandong University, China in 1981 and 1988, respectively, and Ph. D. degree in Physics from University of Cologne, Germany in 1992.

 He was a post-doctoral fellow at Fraunhofer Institute for Applied Optics and Precision Instrument, Jela, Germany in 1993 and a research fellow in Department of Electronic Engineering, City University of Hong Kong from 1994 to 1996. From 1996 to 2013, he worked in Data Storage Institute (DSI), Singapore as a senior scientist and division manager, and led nonvolatile solid-state memory (NVM), artificial cognitive memory (ACM) and optical storage researches. He is the recipient of the National Technology Award 2004 Singapore, the only one awardee that year. He joined THU, China, as a national distinguished professor and director of Optical Memory National Engineering Research Center in 2013. By integrating 7 departments (Precision Instrument, Computer Science, Electrical Engineering, Microelectronics, Medical School, Automatics, Software School), he established CBICR, THU in 2014, and served as the director.
  His  research interests include brain-inspired computing, neuromorphic engineering, NVM, memristor devices, optical data storage, photonics, etc.

 Dr. Shi has published more than 150 papers in prestigious journals including Science,Nature Photonics, Advanced Materials, Physical Review Letters, filed and granted more than 10 patents and conducted more than 60 keynote speech or invited talks at many important conferences during last 10 years. Dr. Shi is a SPIE fellow, and general co-chair of the 9th Asia-Pacific Conference on Near-field Optics2013, IEEE NVMTS 2011-2015, East-West Summit on Nanophotonics and Metal Materials 2009 and ODS009. He is a member of the editorial board of Scientific Reports (Nature Publishing Group).
\end{IEEEbiography}






\end{document}